\documentclass[a4paper,11p]{article}

\pdfoutput=1 

\usepackage{jinstpub}

\usepackage[english]{babel}
\usepackage{units}
\usepackage{graphicx}
\usepackage{caption}

\usepackage{color}
\usepackage{xspace}

\usepackage{caption}
\usepackage{subcaption}

\DeclareMathOperator{\sgn}{sgn}

\definecolor{dark-gray}{gray}{0.3}

\newcommand{\Xmax}{$X_{\rm max}$\xspace}

\title{A high-precision interpolation method for pulsed radio signals from cosmic-ray air showers}

\author[a,b,1]{A.~Corstanje}
\author[a,b]{S.~Buitink}
\author[a]{M.~Desmet}
\author[b,c,d]{H.~Falcke}
\author[c]{B.M.~Hare}
\author[b,d,a]{J.R.~H\"orandel}
\author[f,a]{T.~Huege}
\author[j]{V.~B.~Jhansi}
\author[f]{N.~Karastathis}
\author[a]{G.~K.~Krampah}
\author[a]{P.~Mitra}
\author[b,d]{K.~Mulrey}
\author[g,h]{A.~Nelles}
\author[b]{K.~Nivedita}
\author[a]{H.~Pandya}
\author[e,i]{O.~Scholten}
\author[h]{K.~Terveer}
\author[j]{S.~Thoudam}
\author[k]{G.~Trinh}
\author[c]{S.~ter Veen}

\affiliation[a]{Vrije Universiteit Brussel, Astrophysical Institute, Pleinlaan 2, 1050 Brussels, Belgium}
\affiliation[b]{Department of Astrophysics/IMAPP, Radboud University Nijmegen, P.O. Box 9010, 6500 GL Nijmegen, The Netherlands}
\affiliation[c]{Netherlands Institute for Radio Astronomy (ASTRON), Postbus 2, 7990 AA Dwingeloo, The Netherlands}
\affiliation[d]{Nikhef, Science Park Amsterdam, 1098 XG Amsterdam, The Netherlands}
\affiliation[e]{University of Groningen, Kapteyn Astronomical Institute, Groningen, 9747 AD, Netherlands}
\affiliation[f]{Institut f\"{u}r Astroteilchenphysik, Karlsruhe Institute of Technology (KIT), 
 P.O. Box 3640, 76021, Karlsruhe, Germany}
\affiliation[g]{Deutsches Elektronen-Synchrotron DESY, Platanenallee 6, 15738 Zeuthen, Germany}
\affiliation[h]{ECAP, Friedrich-Alexander-Universit\"{a}t Erlangen-N\"{u}rnberg, 91058 Erlangen, Germany}
\affiliation[i]{Interuniversity Institute for High-Energy, Vrije Universiteit Brussel, 
 Pleinlaan 2, 1050 Brussels, Belgium}
\affiliation[j]{Department of Physics, Khalifa University, P.O.~Box~127788, Abu Dhabi, United Arab Emirates}
\affiliation[k]{Department of Physics, School of Education, Can Tho University Campus II, 
 3/2 Street, Ninh Kieu District, Can Tho City, Vietnam}

\emailAdd{A.Corstanje@astro.ru.nl}

\abstract{
Analysis of radio signals from cosmic-ray induced air showers has been shown to be a reliable method to extract shower parameters such as primary energy and depth of shower maximum.
The required detailed air shower simulations take 1 to 3 days of CPU time per shower for a few hundred antennas.
With nearly $60,000$ antennas envisioned to be used for air shower studies at the Square Kilometre Array (SKA), simulating all of these would come at unreasonable costs.
We present an interpolation algorithm to reconstruct the full pulse time series at any position in the radio footprint, from a set of antennas simulated on a polar grid.
Relying on Fourier series representations and cubic splines, it significantly improves on existing linear methods. 
We show that simulating about 200 antennas is sufficient for high-precision analysis in the SKA era, including e.g.~interferometry which relies on accurate pulse shapes and timings.
We therefore propose the interpolation algorithm and its implementation as a useful extension of radio simulation codes, to limit computational effort while retaining accuracy.
}

\begin{document}
\maketitle
\flushbottom


\section{Introduction}
The radio detection of cosmic rays has advanced considerably over the past decade, with instruments such as LOFAR \cite{vanHaarlem:2013dsa} and AERA \cite{Huege:2019snr} allowing for detailed reconstructions of individual air showers. 
To reconstruct and analyze the properties of the radio signal on the ground, detailed ('microscopic') simulations are typically used, in which individual particles are tracked along their trajectory and interactions in the atmosphere, summing their contribution to the radio signal at several antenna locations on the ground. These simulations are then fitted to measured data e.g.~\cite{Buitink:2014} or are used to develop parameterized reconstruction methods for the signals, e.g.~\cite{Nelles:2014gma, Glaser:2016qso}.  

Simulation programs such as Corsika \cite{Corsika:1998} plus CoREAS \cite{CoREAS:2013} and ZHAireS \cite{ZHAireS:2012} were found to be highly accurate \cite{Gottowik:2017wio}, but are also time consuming in generating showers; for an average detector configuration comprising 150 to 200 antennas, simulating a single air shower takes on the order of 1 to 2 days in a single CPU thread, and even longer at the highest energies. The running time is dominated by computing the radio signal, and scales proportionally with the number of antennas.
A high-precision cosmic-ray mass composition analysis such as done at LOFAR \cite{Corstanje:2021} or at AERA at the Pierre Auger Observatory \cite{Pont:2021} requires accurate reconstructions of the energy and depth of shower maximum, \Xmax, for which Monte Carlo ensembles of about 30 simulated showers per measured air shower have been used. This is a significant computational burden. 

The upcoming Square Kilometre Array observatory (SKA-Low) \cite{Tan:2015} will feature nearly $60,000$ antennas in an inner circle with a radius of $\unit[500]{m}$. Cosmic-ray measurements here will naturally be aimed at a very high precision in characterizing the radio signal from air showers \cite{Huege:2016jvc}.
However, at two orders of magnitude more antennas than LOFAR, simulating the signals at all antenna positions is intractable with the current setup. The method we present aims to reduce the number of antennas to be simulated per shower to around 200 while retaining accuracy. This would keep the computational load for dedicated Monte Carlo ensembles per shower, which would still be needed, about the same as for earlier analyses.

\subsection{Air shower pulse characteristics}\label{sect:characteristics}

\begin{figure}
    \centering
    \includegraphics[width=0.49\textwidth]{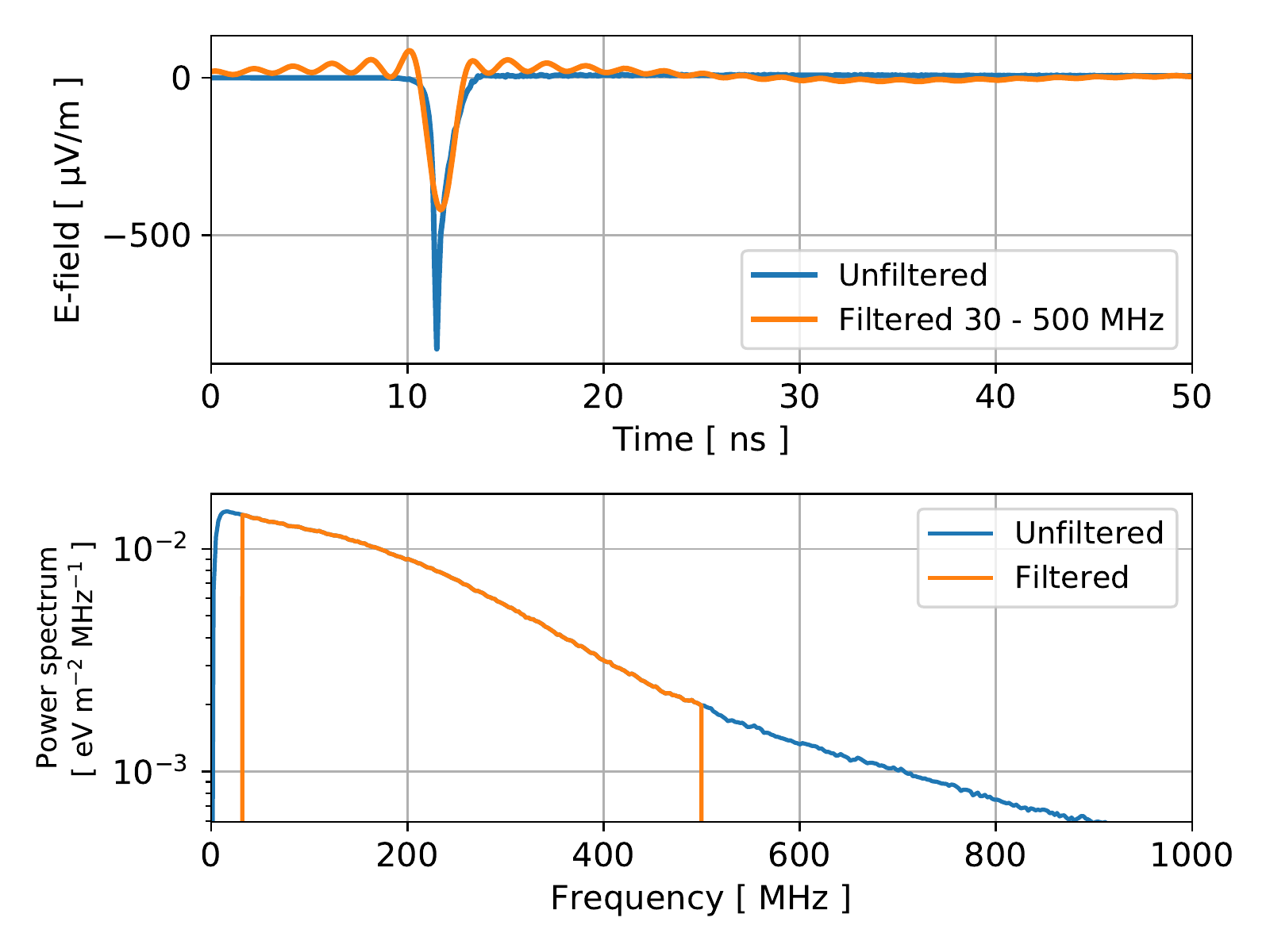}
    \includegraphics[width=0.49\textwidth]{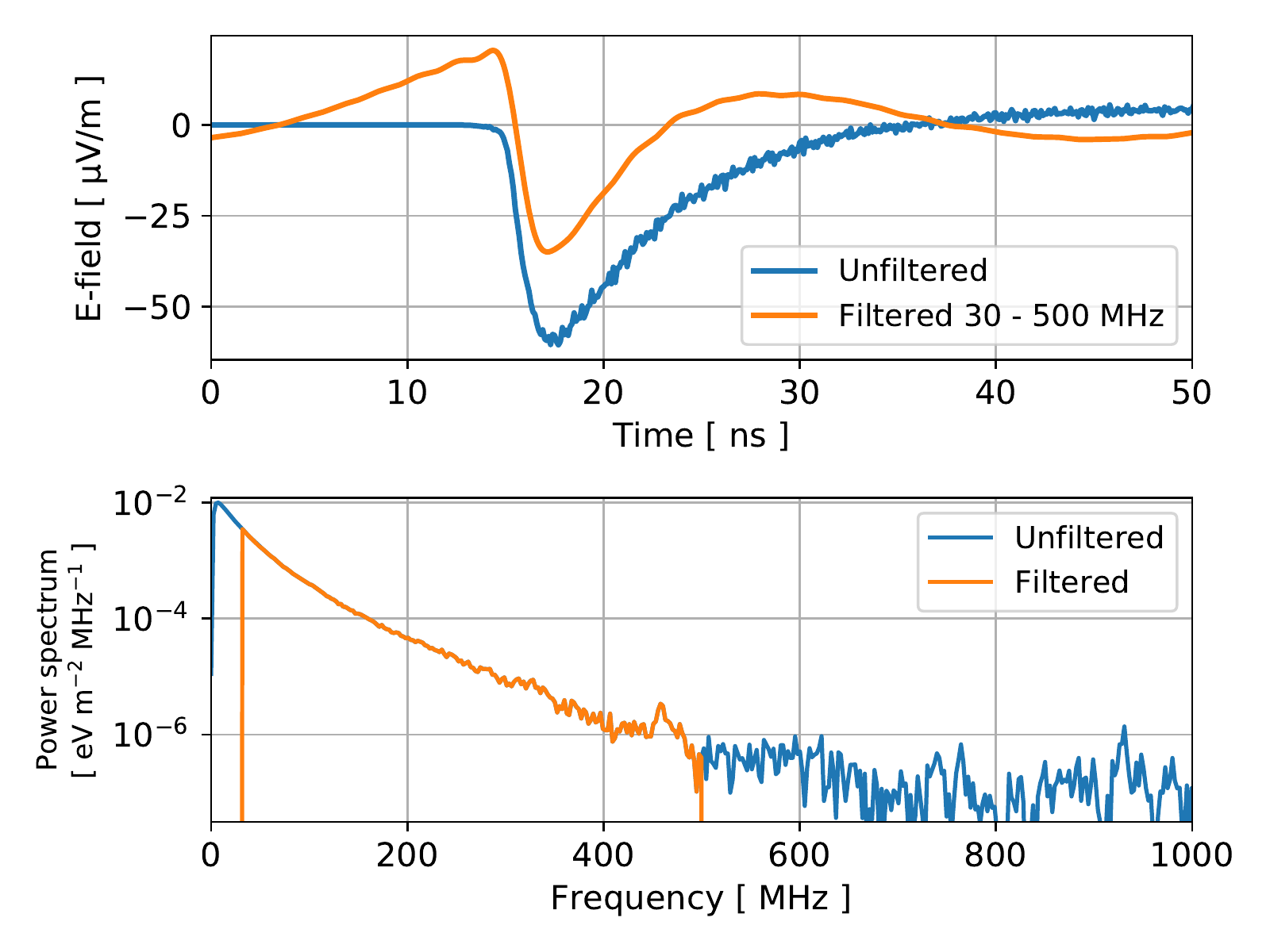}
    \includegraphics[width=0.49\textwidth]{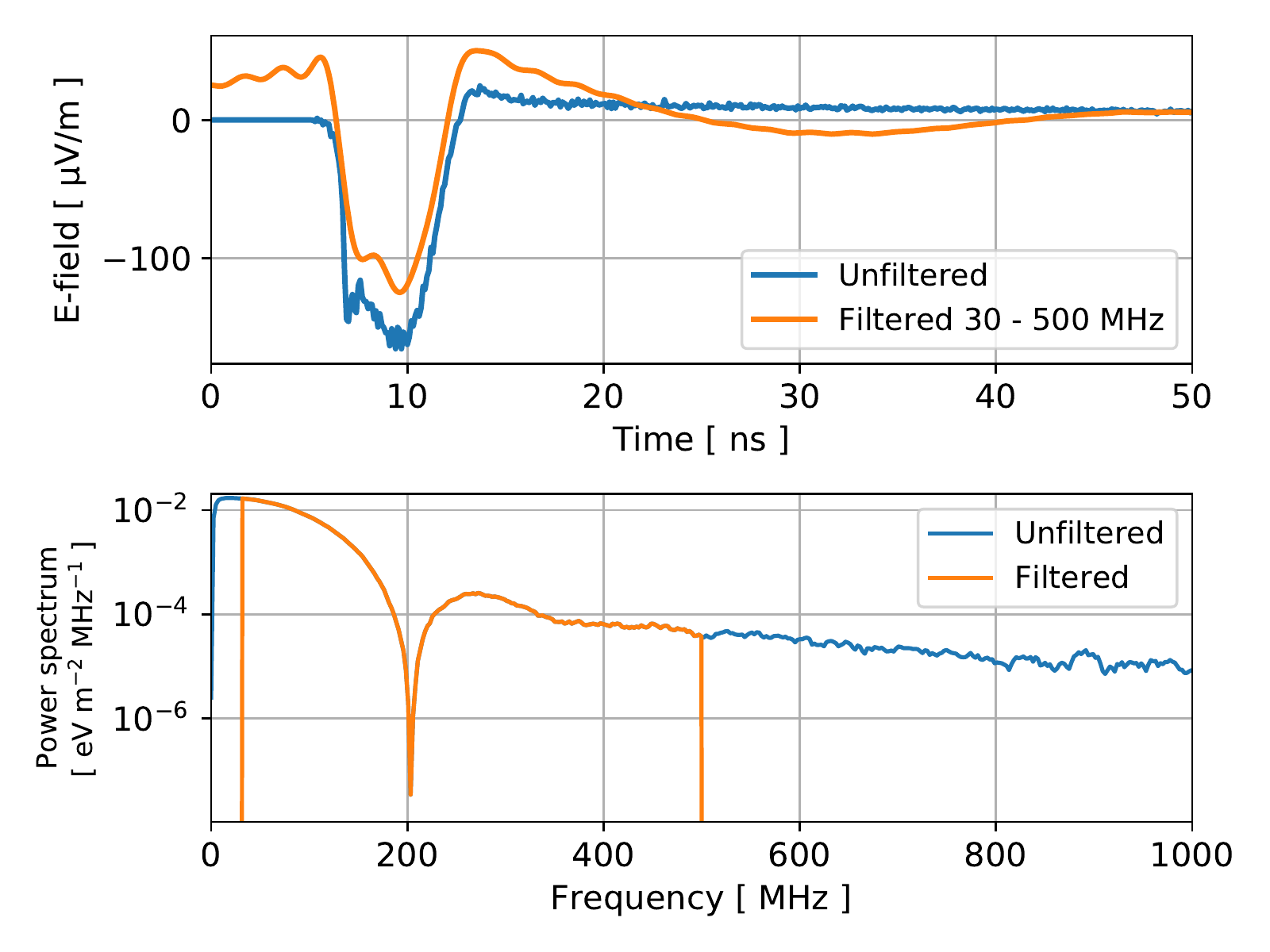}
    \includegraphics[width=0.49\textwidth]{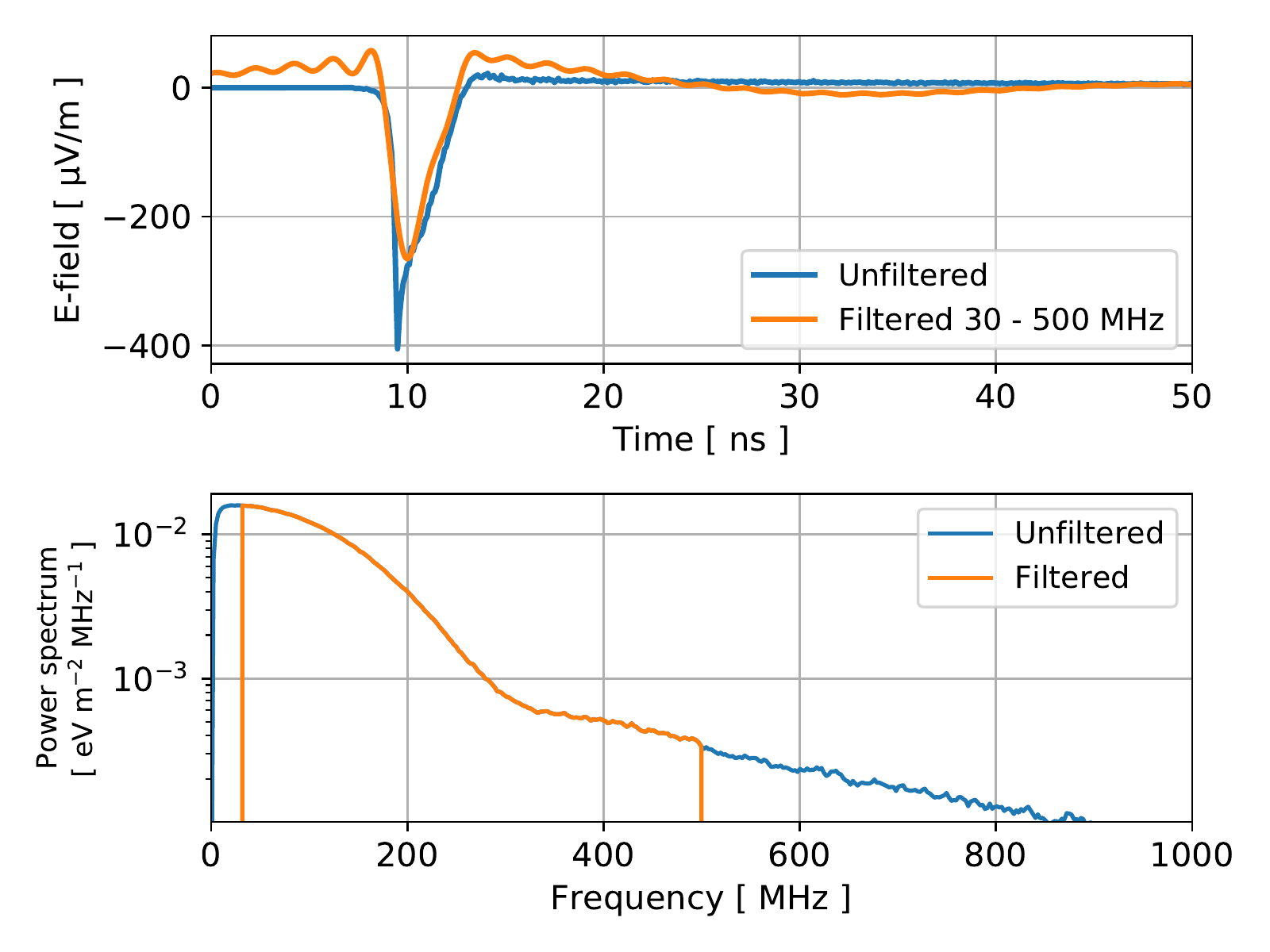}
    \caption{Four panels of different CoREAS pulses in the time-domain (top) and in the frequency domain (power spectrum, bottom). The pulse shape, and thus the frequency content, changes dramatically moving outward in the shower plane. For clarity, these figures only show one component of the electric field, thus no information about the polarization. For phase information confer Fig.~\ref{fig:phasespectrum}.}
    \label{fig:Pulse_didactic}
\end{figure}

Air shower pulses as they arrive at the ground show considerable variations in time-domain pulse shape as well as frequency spectrum (see Fig.~\ref{fig:Pulse_didactic}).

The pulse length matches the time scale or transversal length scale of an air shower; it is intrinsically a bipolar pulse of only a couple of nanoseconds in duration. This is, however, true only for unfiltered pulses. Filtering typically induces ringing, which extends the duration and provides a more complex pulse shape at the receiver.  

Correspondingly, the unfiltered frequency spectrum spans a broad-frequency range, typically peaking at the low MHz frequencies, with an (almost) exponential decay towards higher frequencies. The decay rate changes as function of angular distance to the Cherenkov angle. So if one moves in the shower footprint from the shower axis outwards, the high frequency content first increases until the Cherenkov angle (for LOFAR or SKA typically around 100~m distance) and then decreases again. 

The pulses are strongly polarized. Their electric-field vector is determined by the two main radio emission mechanisms. The dominant geomagnetic emission generates pulses polarized along the $\mathbf{v}\times\mathbf{B}$-axis, with $\mathbf{v}$ the shower axis and $\mathbf{B}$ the magnetic field. The Askaryan effect as secondary emission, also referred to as charge-excess component, alters this dominant signal by up to 20\% by adding a field component pointing in the direction of the shower axis. This is thus different for every observer position in the shower footprint. So, the resulting electric field can either increase or decrease, due to constructive or destructive interference depending on the observer position.

CoREAS, the simulation software used for LOFAR and AERA analyses, calculates these pulses to a high accuracy. A number of example pulses are shown in Fig.~\ref{fig:Pulse_didactic} to illustrate the complexity of the interpolation problem. These are simulated pulses, shown unfiltered as well as bandpass-filtered using an `ideal' FFT filter. In a real antenna and signal chain, the time series will look different, due to additional filtering and the antenna characteristics. Nevertheless, the simulated signals represent the input to the instrument-dependent signal chain, and as such, we keep the discussion and analysis instrument-independent here.

Clean pulses as shown in the upper left panel are relatively easy to interpolate, they can almost be approximated by a straight line in the logarithmic power spectrum. However, as seen in the bottom left panel, the power spectrum may also have nontrivial features at some positions in the footprint. In this case, a zero is seen near $\unit[200]{MHz}$; the phase spectrum makes a jump of $\pi$ there, corresponding to a minus sign. Thus, the energy spectrum goes to zero there, switches sign, and comes up again to about $1/100$ of the peak level, towards higher frequencies.
This effect is seen in many showers inside the Cherenkov-ring, at distances around $10$ to $\unit[40]{m}$ from the shower core, which makes it a complex region to interpolate pulses, in particular when subtle effects need to be captured. 

In addition to what we believe are true physical peculiarities, the simulated CoREAS pulses also contain some artifacts that one needs to be aware of, when working with them. CORSIKA employs thinning to be computationally efficient. In this thinning, particles are grouped, which introduces artificial coherence for the radio simulations and leads to what is typically referred to as \emph{thinning noise}. This is most easily identified as an additional noise contribution that dominates at high(er) frequencies, where the true air shower signal drops off. The noise scales with shower energy, which makes it easily identifiable in an ensemble of air showers. In single showers, special care has therefore to be taken to not include this artificial power contribution in an analysis. 

Additional examples of simulated pulses, along with the interpolated pulses using the presented method, are shown in the Appendix.

\subsection{Signal processing}
In general, a Fourier spectrum of a signal consists of complex values which can be separated into an amplitude or power spectrum (see Fig.~\ref{fig:Pulse_didactic}), and a phase spectrum. 
A pure, bandpass-filtered impulse has a simple, linear phase spectrum as a function of frequency $\nu$, 
\begin{equation}\label{eq:linear_phase}
    \phi(\nu) = \phi_0 + 2\pi\, \nu\, \Delta t  \pmod{2\pi},
\end{equation}
with a slope corresponding to the arrival time $\Delta t$, and an intercept $\phi_0$.
This implies that adding a linear function to $\phi(\nu)$ directly proportional to $\nu$ shifts the signal in time, without changing its shape.

The phase constant $\phi_0$, which we could also call {\it Hilbert} phase, is zero or $\pi$ for a symmetric pulse, i.e.,~an even function, and $\pm \pi/2$ for an antisymmetric or odd function. 
Pulses with different $\phi_0$ but otherwise the same phase spectrum, share the same {\it Hilbert envelope}, which is a non-negative function tracing the `envelope' of the signal, not disturbed by the natural oscillations.
The definition of the Hilbert envelope, here denoted $A(t)$, is
\begin{equation}
 A(t) = \sqrt{x^2(t) + \hat{x}^2(t)},
\end{equation}
where $\hat{x}(t)$ is the Hilbert transform of the signal $x(t)$. Typically the signal is upsampled to determine the maximum to higher precision than the original sampling rate. The Hilbert transform itself is defined by 
\begin{equation}
\mathcal{F}\left[\hat{x}(t)\right](\nu) = -i\; \sgn (\nu)\; \mathcal{F}\left[x(t)\right](\nu),
\end{equation}
where $\mathcal{F}$ is again the Fourier transform. Effectively it applies a $90^{\circ}$ phase shift to each frequency.
For $\phi_0 = 0$ or $\pi$, the maximum of the Hilbert envelope coincides with the maximum of the signal itself (in our case, the electric field) in the time domain.

The radio air shower community has converged on using energy fluence defined as
\begin{equation}
F = \frac{1}{Z_0} \int_{-T}^{T} E(t)^2\,dt,
\label{eq-fluence}
\end{equation}
with electric field magnitude $E$ and vacuum impedance $Z_0$, for most physics analyses. It describes the energy per unit area in a pulse (see e.g.~\cite{PAO:2016}), which is the time integral of the instantaneous Poynting flux over a certain frequency range of interest, in a time window (taken as $-T$ to $+T$) around the maximum of the pulse, here defined at $t=0$. It is typically expressed in units of $\unit{eV/m^2}$.
This provides a quantity with a physical meaning, that is not as dependent on instrument characteristics as for instance amplitude. In simulations, the integration is usually along the entire waveform; in a realistic measurement setting one may also take the fluence in a narrow time window around the maximum of the signal \cite{Schellart:2013}, capturing a known fraction of the pulse energy to get a reasonable signal-to-noise ratio.

The pulse interpolation introduced here is targeted at providing quantities typically used in radio analyses. They include the electric field as function of time, as well as the arrival time $\Delta t$ and phase constant $\phi_0$ (Eq.~\ref{eq:linear_phase}), amplitude, and energy fluence.

\subsection{Previous work}

A natural way to deal with the computational overhead exists in the form of `macroscopic' simulations making use of parametrizations of particle densities etc.\ instead of tracking individual particles \cite{Scholten:2018,deVries:2013}. As typical running times are less than a minute, these are useful for e.g.~initial estimates of shower properties. Also, they offer a complementary way of describing the air shower process, from which one can gain insight in emission mechanisms (e.g., the geomagnetic and charge-excess contributions).
For high-precision work however, the systematic uncertainties introduced by the parametrizations are generally either too large or insufficiently known. In the code for macroscopic calculations, MGMR~\cite{Code_MGMR}, a pulse interpolation is used based on a combination of interpolating a time shift and the Fourier transforms of the time-shifted pulses which served as an inspiration for the approach presented in the present work.

Other approaches center around creating a library of pre-computed simulations instead of having dedicated simulations per measurement \cite{Zilles:2020}, or on reducing the running time of simulations by having a limited number of antennas and interpolating in between. Such an interpolation offers a parameter-free description of the radio signals along the footprint, based on the signals at the simulated positions.

In this paper, we focus on the latter approach, noting that reducing running time by interpolation would be useful for creating libraries as well.
We present an interpolation method for signal properties such as amplitude and fluence, as well as for the full pulse shape and spectrum. 
While linear interpolation methods exist and work reasonably well for some purposes \cite{Tueros:2021}, our method goes beyond linear interpolation, aimed at high-precision work in present and future observatories such as LOFAR, AERA, and SKA. It makes use of the natural geometry of the air shower signal as described on a radial (polar) grid in the shower plane.

For measurements at LOFAR, using the low-frequency range of 30~to~$\unit[80]{MHz}$, a method based on radial basis functions from the library Scipy \cite{Scipy:2020} was found sufficient for pulse energy and amplitude \cite{Buitink:2014}; a full pulse signal interpolation was never pursued.
However, when working in a wider frequency range such as 50~to~$\unit[350]{MHz}$ at SKA, the previous method as well as available alternatives were found inadequate (see also the Appendix). This is discussed in some more detail in the Appendix.
In the considerably wider frequency range, making use of the information in the frequency spectrum of the signal is expected to be useful for characterizing the air showers \cite{Huege:2016jvc,Barwick:2016mxm}.
Furthermore, approaches based on interferometry are being explored \cite{Schoorlemmer:2020,Schluter:2021egm}, for which both the pulse shape and an accurate arrival time are needed on arbitrary positions in the footprint. This shows the need for an accurate full-signal interpolation.

This paper is structured as follows. In Sect.~\ref{sect:method}, the interpolation method is described, first for single-value observables such as pulse energy fluence or amplitude per antenna, followed by the method for interpolating the full pulse shape, spectrum, and arrival time.
The performance is evaluated in Sect.~\ref{sect:results}, using simulations with 208 antennas on a radial grid, plus 250 `test' antennas at random positions in the shower plane.

\section{Pulse interpolation method}\label{sect:method}
The explanation of the method is split into two parts: one for single-valued observables such as the total radio energy footprint, and one for the pulse signal shape. The code for this interpolation has been implemented in Python, using standard packages such as Numpy \cite{Numpy:2020} and Scipy \cite{Scipy:2020} and is available on github\footnote{\url{https://github.com/nu-radio/cr-pulse-interpolator}} \cite{github_nuradio:2023}.

With CoREAS, we simulate the radio signal as it reaches the ground plane, on a number of pre-defined antenna positions.
For practical purposes it has been found helpful to arrange these observer positions as such that they form a radial grid in the shower plane, which has the incoming direction $\mathbf{v}$ as normal vector, with the geomagnetic field $\mathbf{B}$ being the other defining axis \cite{Buitink:2014}. This set-up is widely used throughout the radio community and will be the basis of our interpolation method.  
It is also called `star-shaped pattern', the rationale behind this being illustrated by the white dots in Fig.~\ref{fig:radialgrid}.

\subsection{Interpolation of the pulse energy footprint}
\label{sect:energy_interpolation}

\begin{figure}
	\centering
	\includegraphics[width=0.70\textwidth]{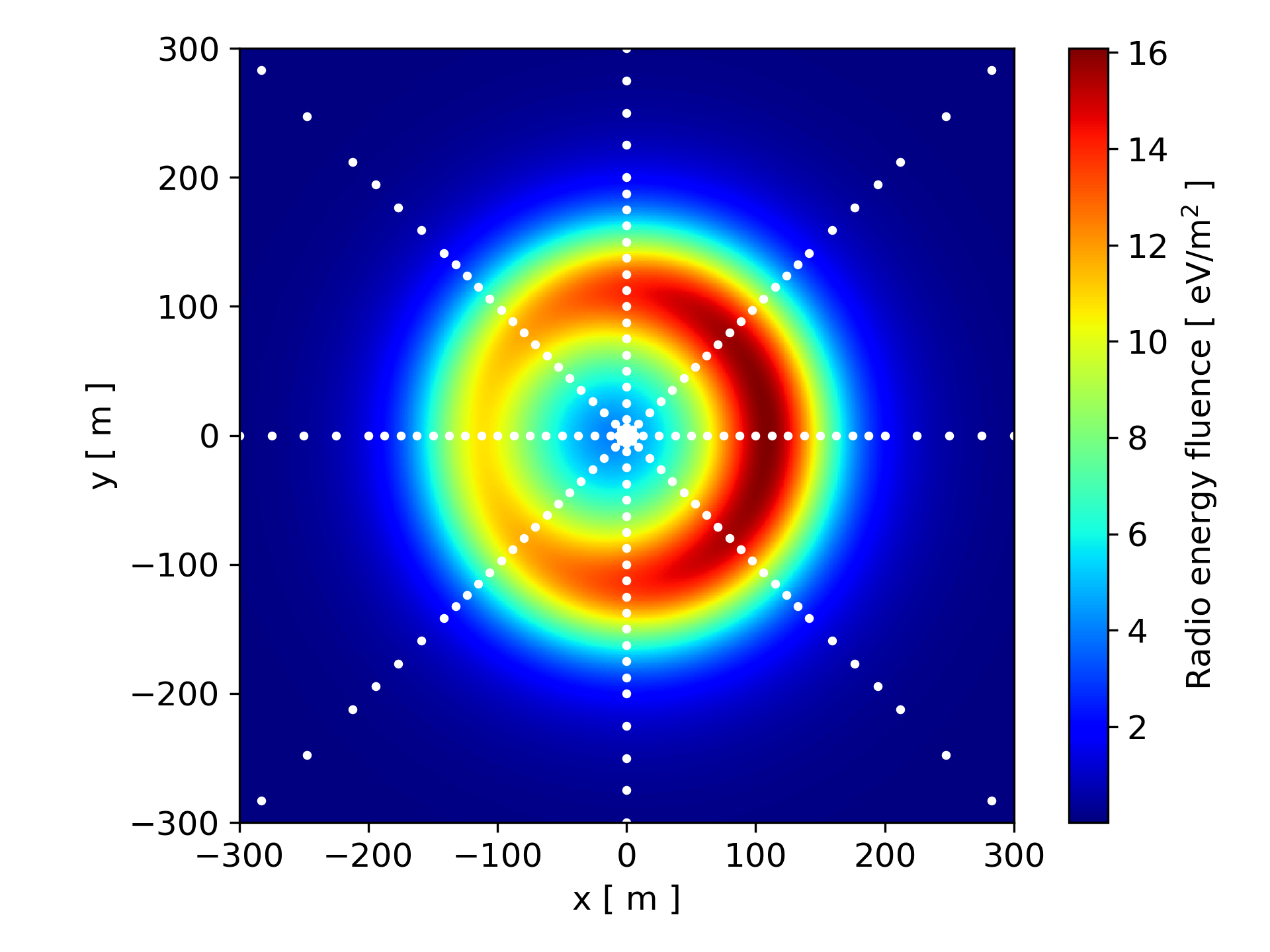}
	\caption{An example footprint of radio energy in the frequency range $30$ to $\unit[500]{MHz}$. It is given in shower-plane coordinates, with $\mathbf{x}$ pointing along $\mathbf{v}\times \mathbf{B}$. The simulated antenna positions on a radial grid are shown as white dots.}
	\label{fig:radialgrid}
\end{figure}

As a first interpolation quantity, we use the \emph{radio energy fluence} as the total energy in the simulated time series signal, summed over all three polarization components, see also Eq.\ref{eq-fluence}. 
For the interpolation, we make use of the geometry of the dominant contributions to the radio signal, being the geomagnetic and charge excess components (see e.g.~\cite{Huege_review:2016} for a recent review). 
At a fixed radius, the amplitudes vary with angular position $\theta$ around the shower axis, expressing positions in the (2D) shower plane in polar coordinates ($r$, $\theta$).
Variations are proportional to $\cos(\theta)$ for the charge excess component, while the geomagnetic part is circular symmetric.

This motivates the use of a Fourier series along the angular direction, i.e.\ circles at constant distance from the axis, to describe variations.
The Fourier series is a complete representation of a periodic signal measured at finite, equidistant sampling intervals, provided the sampling is dense enough to describe the highest (spatial) frequencies, given by the Nyquist criterion. The assumption that 8 angular points suffice (i.e.~8 `arms' in the star-shape pattern) is tested below, in Sect.~\ref{sect:angular_modes}, Fig.~\ref{fig:angular_dependence}.

The interpolation method to obtain a single-valued pulse property at a particular position is set up as follows:
\begin{enumerate}
\item At a fixed radius, take the antenna positions along a circle, and express the variation of the pulse energy along the circle as a Fourier series. The Fourier components are given by a Fast Fourier Transform (FFT) over the antenna positions (in case of Fig.~\ref{fig:radialgrid}, 8 antennas).
\item For convenience, express the Fourier components in (real) cosine and sine amplitudes $c_k(r_i)$ and $s_k(r_i)$ at each simulated radius $r_i$.
\item For the radial dependence, interpolate each of the Fourier components radially using cubic splines, or another better-than-linear interpolation scheme.
\end{enumerate}

The interpolated estimate of the footprint energy $f(r, \theta)$ at a given position is then expressed as
\begin{equation}\label{eq:interpolation_def}
\hat{f}(r, \theta) = \sum_{k=0}^{n/2} c_k(r)\cos(k\,\theta) + s_k(r)\sin(k\, \theta),
\end{equation}
and one evaluates the interpolated value at $(r, \theta)$ from this sum explicitly, after obtaining the angular Fourier components $c_k(r)$, and $s_k(r)$ from one-dimensional radial interpolation of $c_k(r_i)$ and $s_k(r_i)$ taken at the radii $r_i$ of the simulated antennas.

\subsection{Interpolation of the pulse time series signal}\label{sect:method_pulse}
To obtain a good approximation of the electric field as function of time $E(t;\,x,y)$ in between simulated positions $(x_i, y_i)$, a simple linear combination of nearby signals is not appropriate. This stems from the pulse arrival times, the polarization, as well as the pulse shape, that change smoothly but non-linearly along the footprint.
So for a full pulse interpolation, we interpolate in the spectral domain, writing the full spectrum at each position $(x_i, y_i)$ as
\begin{equation}
F(\nu) = \mathcal{F} (E(t)) = \left|F(\nu)\right| \exp\left(\mathrm{i}\,\phi(\nu)\right),
\end{equation}
with $\nu$ denoting frequency, separating the absolute (amplitude) spectrum from the phase spectrum $\phi(\nu)$.
For a fixed frequency channel $\nu$ in the spectrum, the amplitude is a positive real number. Thus, having one real number per antenna position for each frequency $\nu$, these can be successfully interpolated by the algorithm described in the previous section, \ref{sect:energy_interpolation}.

The phase spectrum is more challenging, as it consists of phase factors $\exp(\mathrm{i}\,\phi)$ which would identify $\phi$ with $\phi + 2\,\pi\,k$ for integer $k$. This poses difficulties for any straightforward interpolation method, as it is sometimes unclear in which period-window the intermediate values should lie. There are options to do this, one of which is through directly interpolating $\phi$ and working out which of the $2\pi$ multiples to interpolate between. This is known as {\it phase unwrapping}. For the most generic case of 2D phase data this is a notably difficult problem as it is NP-complete \cite{Chen:2001}, and one would have to hope that the phases in the simulations at hand allow for an easier and correct solution also in the presence of (phase) noise.
An alternative is to consider directly interpolating the phase vector $\exp(\mathrm{i}\, \phi)$.
We found neither to work very well, in part because the phases may differ quite a bit between neighboring antennas, and also because the phases become unstable when the simulated signals get weak.

Therefore, from the phase spectrum we first examine some non-periodic properties that can be extracted. The idea is that non-periodic quantities are readily interpolated using the single-value interpolation from the previous section, and then be used to reconstruct the phase spectrum again.

For a short, coherent pulse, the phase spectrum is mainly determined by a linear function of frequency (see Sect.\ref{sect:characteristics}, and Eq.~\ref{eq:linear_phase}). When higher-order terms, i.e.\ {\it dispersion} is small, it can be neglected over a limited frequency window.

The linear components amount to $\phi_0$, the intercept, and $\Delta t$, the slope defining the arrival time.
The maximum of the Hilbert envelope defines $\Delta t$.
We obtain the phase constant $\phi_0$, by summing over the complex spectrum after taking out the time shift $\Delta t$, and taking the angle:
\begin{equation}
\phi_0 = \arg \,\sum_{\nu} F(\nu)
\end{equation}
Obtaining $\phi_0$ and $\Delta t$ in a low-frequency range of e.g.~$30$ to $\unit[80]{MHz}$ is most stable, as the signal gets weaker towards higher frequencies and larger distances from the shower core.
We now have a zeroth and first-order phase spectrum taken out, from which we obtain a value $\phi_0$ and $\Delta t$ at each antenna position.
The arrival time $\Delta t$ is interpolated using the method in Sect.~\ref{sect:energy_interpolation}; for $\phi_0$, which varies along a simulated footprint, we unwrap the values by taking for neighboring values the multiple of $2\pi$ that minimizes the difference. Unwrapping first along the radial direction, then along circles was found to work reliably in this case, as $\phi_0$ varies only slowly with position.

\begin{figure}
	\centering
	\includegraphics[width=0.70\textwidth]{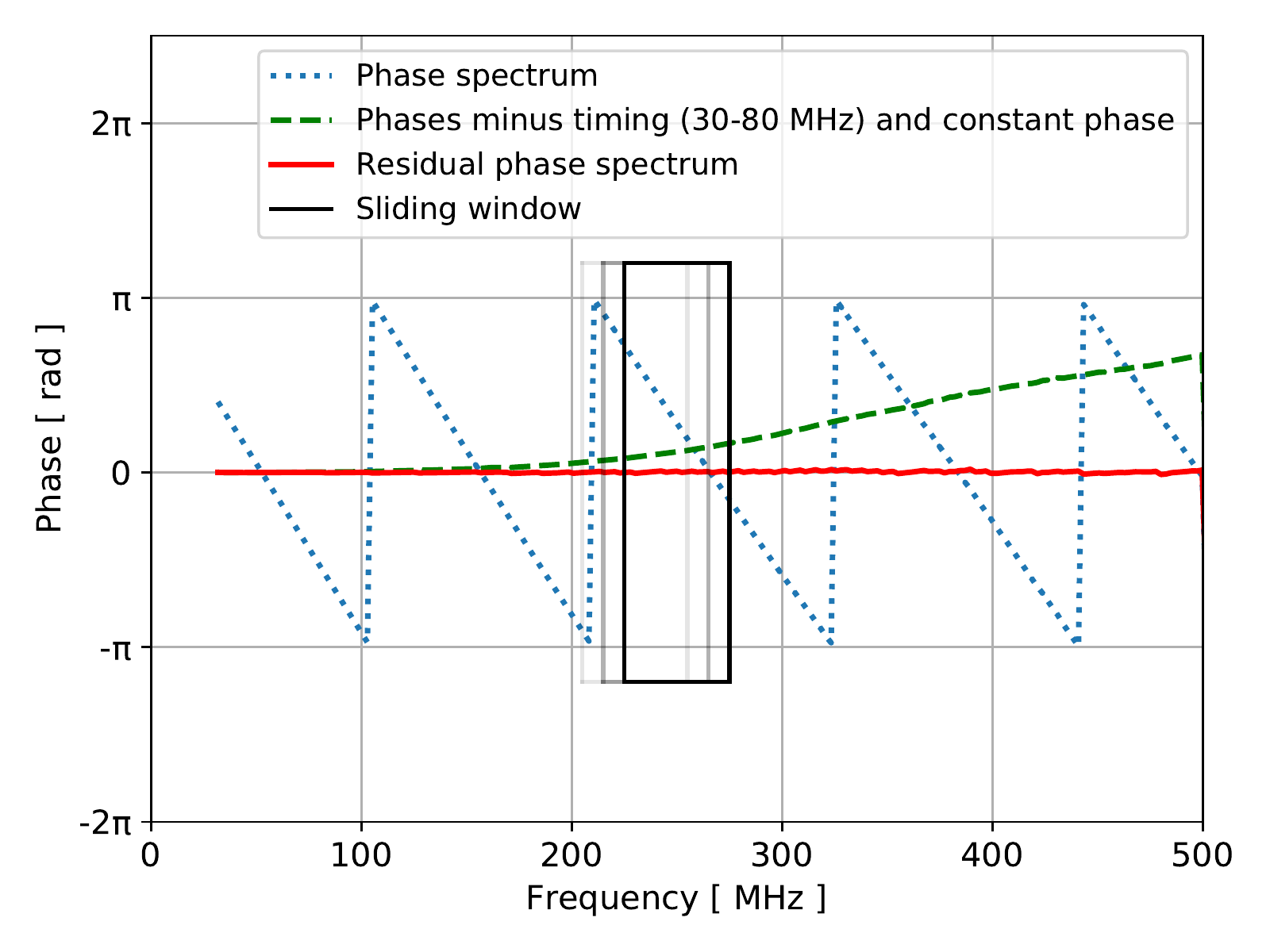}
	\caption{An example of a phase spectrum from the signal at $x=\unit[100]{m}$ and $y=0$, close to the maximum intensity in the footprint. Shown are the phase spectrum stemming from CoREAS, the remaining  phase spectrum after subtracting a linear contribution due to timing taken between 30 and $\unit[80]{MHz}$, and the near-zero residuals after subtracting contributions from arrival times as a function of frequency window.}
	\label{fig:phasespectrum}
\end{figure}

What remains unaccounted in the phase spectrum are the higher-order terms, i.e.,~the {\it dispersion} $\phi(\nu) \sim \nu^2$ and higher.
An example of this is shown as the green line in Fig.~\ref{fig:phasespectrum}.
It shows a slow variation with frequency, where the part from $\nu > \unit[300]{MHz}$ has again a roughly linear slope, indicating that the high-frequency part of the pulse arrives slightly earlier than the low-frequency part. We have found two approaches for dealing with this remaining phase spectrum, avoiding periodicity ambiguities whenever possible. They were found to perform about identically well on our specific test set and antenna layout; the user can choose between the two when testing the method for a different use case.

The first, and simplest approach is to express the phase values in the residual phase spectra as complex unit vectors, i.e.\ {\it phasors} defined as $\exp(\mathrm{i}\, \phi(\nu))$. The real and imaginary part are interpolated using the Fourier method, and one takes the angle (argument) of the interpolated complex number to obtain an approximation to the phase.
This is not exactly the same as interpolating pure phases, as a linear function of two complex numbers does not yield a linear progression from one phase to another, nor does it yield a linear progression between the amplitudes of the two numbers.
Nevertheless, given its simplicity it performs well in practice. A couple of variants can be envisioned, for instance interpolating complex amplitudes (i.e.\ amplitude and phase together) to extract both amplitude and phase, or the same to extract only the phase. These were found to yield slightly worse results than using phasors.

The second approach is to apply essentially the above procedure for higher frequencies as well, extracting linear phase spectrum quantities in a sliding window of $\unit[50]{MHz}$ width, starting at $\unit[30]{MHz}$ to $\unit[80]{MHz}$, then $\unit[32]{MHz}$ to $\unit[82]{MHz}$, etc. This approximates the signal in each $\unit[50]{MHz}$ window as a dispersionless pulse. Using a sliding window instead of separate bands ensures continuity along the spectrum.

The arrival time, either computed as the positive maximum in the trace, or from the Hilbert envelope maximum and the remaining constant phase term, closely approximates the phase value in the middle of the window. Hence, the arrival time in each of the frequency windows maps to the phase in the middle of the window.
Obtaining a timing has the strong advantage of returning an unambiguous time value, avoiding the $2\pi$ periodicity of phases.
Interpolating the `timing spectrum' thus found, and converting it back to phases, delivers the interpolated signal. 

However, a problem with pulse timing is that when our simulated signals become very weak, they no longer form a single, coherent pulse but become `noise'. Timing values would become very inaccurate there, which raises the need for a stopping criterion.
The amplitude spectrum of the signals fall off rapidly for locations far from the shower core, hence there is a `cutoff frequency' beyond which the signal cannot be accurately timed or measured.

Beyond the frequency cutoff, which we define below, we keep the arrival time values constant as a best estimate.

We define a \emph{degree of coherency} by considering the complex spectrum after time-shifting the signal maximum to $t=0$ and accounting for the constant phase. We use
\begin{equation}\label{eq:cutoff}
C = \frac{\left| \sum_\nu F(\nu) \right|}{\sum_\nu \left|F(\nu)\right|},
\end{equation}
to obtain a degree of coherency $C$ that measures how well the phases align (the numerator) as a fraction of maximum alignment (denominator, summing absolute values). Its value is between 0 and 1, and will generally drop with increasing frequency.

We define a high-frequency cutoff by setting a threshold value of e.g.~$C > 0.9$, when applying Eq.~\ref{eq:cutoff} to a sliding window of $\unit[50]{MHz}$ width. This threshold value is a tunable parameter, for which $0.9$ has been found to work reliably.
Thus, the method gives an estimate up to what frequency it would be reliable, at any position in the footprint.
The user has the option to low-pass filter the interpolated signal up to this cutoff, if a conservative approach is desired. 

Having discussed the ingredients, let us summarize the time series interpolation algorithm as follows:

\begin{enumerate}
\item Take the FFT of the signal of all antennas and polarizations and split it in absolute-amplitude and phase spectrum.
\item Interpolate the amplitude spectrum using Fourier interpolation (the algorithm described in Sect.~\ref{sect:energy_interpolation}) for each frequency channel in the spectrum.
\item Determine the generic pulse arrival time in all signals, defined by the Hilbert envelope maximum of signals upsampled by a factor 4 to 8, filtered to $30$ to $\unit[80]{MHz}$.
\item Determine the phase constant (`Hilbert phase') and account for phase constant and arrival times, i.e.~taking the corresponding linear function out of the phase spectra. The residual phase spectrum will look like the green line in Fig.~\ref{fig:phasespectrum}, and is essentially zero at low frequencies.
\item {\bf Method (1)}: 
\begin{enumerate}
\item[5.1]
To account for the residual phase spectrum, express the phases (after subtracting the first-order contribution) as complex phasors $\exp({\mathrm{i}\, \phi})$ and interpolate their real and imaginary part. 
\end{enumerate}
{\bf Method (2):}
\begin{enumerate}
\item[5.1] Repeat the above steps and determine pulse timing and phase constant in a sliding frequency window. Obtain a timing value at frequency $\nu$ from band-pass filtering in a window from $\nu - 25$ to $\nu + \unit[25]{MHz}$.
\item[5.2] Determine level of coherency versus frequency in all signals, and for each antenna signal, a high-frequency cutoff where coherency $C$ drops below a threshold level (default, $C=0.9$).
\item[5.3] To prevent meaningless arrival times at high frequencies or weak signals, keep timing values constant beyond the cutoff frequency at each antenna.
\item[5.4] Account for timing as a function of frequency, taking it out of the phase spectrum. The remaining phase spectrum should be close to zero (see the red line in Fig.~\ref{fig:phasespectrum}).
\end{enumerate}
\item Interpolate the generic arrival times, the phase constant, and if applicable the timings as function of frequency, each using the algorithm described in Sect.~\ref{sect:energy_interpolation}
\item For the timing-based method, the remainder of the phase spectrum, which should be almost but not exactly zero, is not interpolated using the Fourier method. Instead, we take the nearest-neighbor values at each position. This is done mainly to ensure the algorithm returns the exact values at the simulated antenna locations that were given as input (as a pure interpolation method should do).
\end{enumerate}

After having obtained the interpolated quantities by using this procedure, the amplitude and phase spectrum of the signal in each polarization are reconstructed at any given arbitrary position $(x, y)$.

\section{Performance of the pulse interpolation}
\label{sect:results}

We have tested the algorithms using a set of air showers simulated with CoREAS.
After presenting the details of the shower ensemble in Sect.~\ref{sect:parameters}, we show some of the inner workings of the Fourier interpolation of total pulse energy fluence along the footprint, in Sect.~\ref{sect:angular_modes}. Then, in Sect.~\ref{sect:results_energy} we discuss the interpolation accuracy of the energy fluence and pulse amplitude, followed by the results of the full pulse interpolation in Sect.~\ref{sect:results_pulse}.

\subsection{Parameters and settings for the simulated shower ensemble}\label{sect:parameters}
We have simulated a total of 84 showers, with $\unit[10^{17}]{eV}$ protons as primary particle.
We have set the altitude of the observer positions to $\unit[460]{m}$ above sea level, and the magnetic field vector to the value at the SKA-Low site in Australia \cite{wmm:2020}, which is $\unit[55.6]{\mu T}$ at an inclination of $\unit[60.24]{^\circ}$.
The showers were simulated using a discrete set of arrival directions. They comprise zenith angles of 15$^\circ$, 30$^\circ$, 45$^\circ$, and 80$^\circ$, and azimuth angles of 0$^\circ$, 15$^\circ$, 30$^\circ$, 45$^\circ$, 90$^\circ$, 135$^\circ$, and 180$^\circ$, as measured clockwise from North for the incoming direction.
By using more arrival directions in azimuth near North, we can better probe the accuracy of the interpolation at low geomagnetic angles (in the Southern hemisphere), where the geomagnetic effect dominates less and the polarization shows stronger variations. 
For each direction we simulated 3 showers. 

We use 208 antennas positioned at ground level, such that they form a radial grid when projected onto the shower plane. The radial spacing is $\unit[12.5]{m}$ up to $\unit[200]{m}$ from the shower axis, then increasing to 25, 50 and $\unit[100]{m}$ up to $\unit[500]{m}$ distance. Three extra radial positions were used in the inner $\unit[12.5]{m}$ from the axis, to have extra coverage there. This selection of antenna positions has proven useful for LOFAR and SKA data analysis. 
For inclined showers, this layout is scaled by multiplying all positions by a factor $4$.
This set of antennas is used to interpolate the signals. 

\begin{figure}
	\centering
	\includegraphics[width=0.60\textwidth]{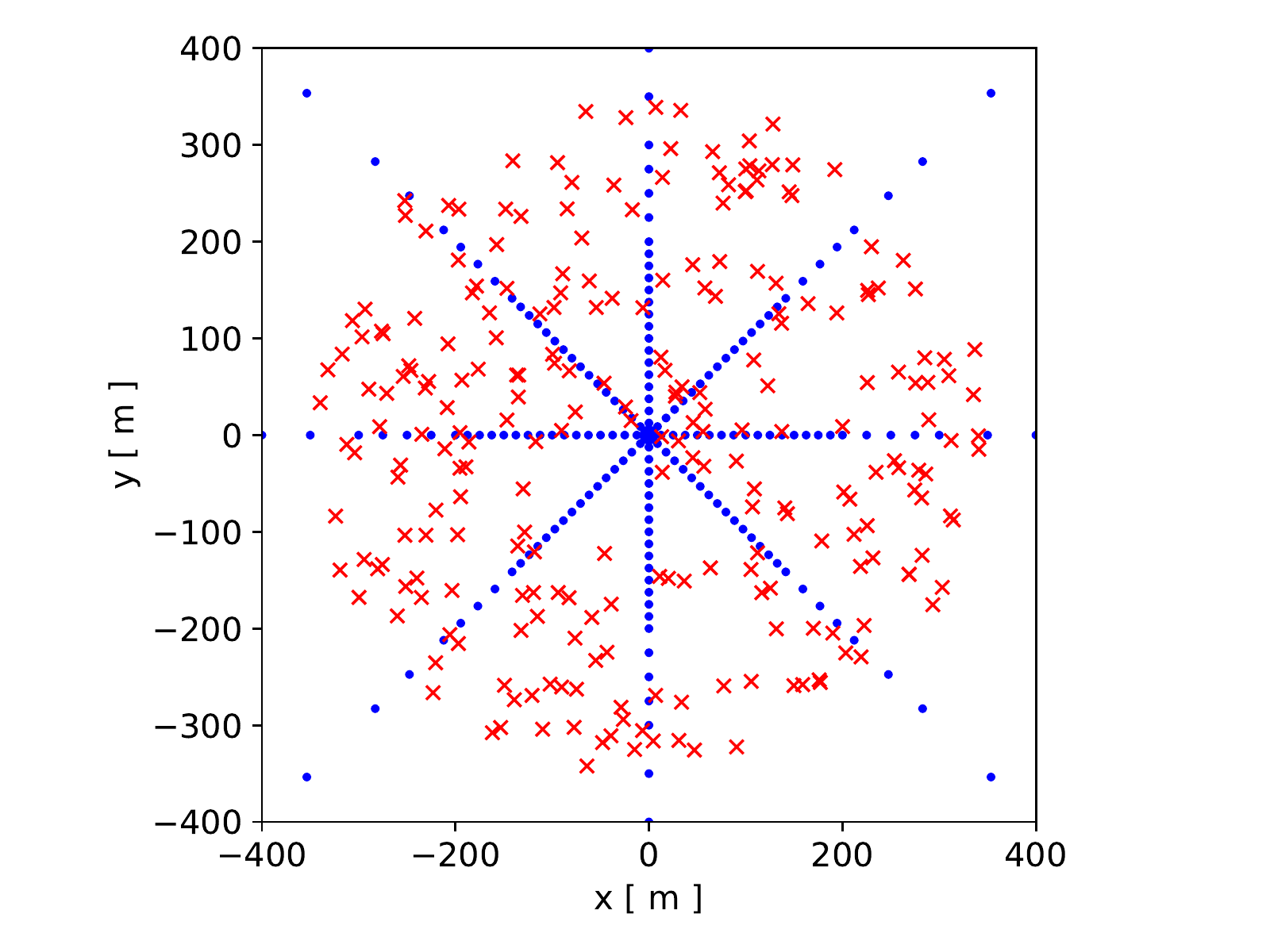}
	\caption{The simulated antenna layout for testing the interpolation. Blue dots indicate the radial grid positions, and red crosses denote the random testing positions.}
	\label{fig:antenna_layout}
\end{figure}

To test the interpolation, we have simulated $250$ additional antennas per shower, positioned randomly within a radius of $\unit[350]{m}$ in the shower plane. One of these layouts is depicted in Fig.~\ref{fig:antenna_layout}; random positions are different for each simulated shower geometry.
Hence, over the full ensemble we have $21,000$ antenna positions on which to test the interpolation method, $7,000$ of which are unique random positions.

Showers were simulated with CORSIKA v7.7410 including CoREAS, with the QGSJetII-04 hadronic interaction model \cite{Ostapchenko:2013} and the UrQMD model for low-energy interactions. 
The time resolution is $\unit[0.1]{ns}$ in a time window of $\unit[408.2]{ns}$. 
Notably, we set the thinning threshold to $10^{-7}$ of the primary energy, with maximum particle weights at $10^{-7}$ of the primary energy in GeV. Typically, thinning thresholds are set to $10^{-6}$ (e.g., for LOFAR data analysis \cite{Corstanje:2021,Buitink:2016}) or even $10^{-5}$ depending on the use case. By reducing thinning, the interpolation accuracy will better reflect the quality of interpolation of the underlying physics, with thinning artifacts being suppressed. It does lead to a factor 5 to 10 longer computation time, so in the present version of the simulation codes, it is not a sustainable setting for large-scale data analysis. For a selection of 15 showers, we have also simulated showers at thinning threshold $10^{-6}$, to evaluate the difference in interpolation results.
In addition, to demonstrate the angular Fourier modes, we have simulated a dedicated shower, featuring two rings of 48~antennas apart from a standard radial grid, at $10^{-6}$ thinning level.

As input for the interpolator of the full signal we use the electric field signals that are provided by CoREAS, on the 208 standard antenna positions. The electric field traces comprising 3 polarizations are transformed to `on-sky' polarizations. These are two orthogonal polarizations in the shower plane, i.e., with the incoming direction $\mathbf{v}$ as normal vector.
The polarization along the incoming direction is usually neglected, as the signal is (close to) zero there. Hence, the two polarizations suffice for a good description, and are a standard tool used when applying antenna models.
The polarizations are taken (or rotated) such that they do not align with $\mathbf{v} \times \mathbf{B}$ and $\mathbf{v} \times (\mathbf{v} \times \mathbf{B})$, to avoid having zeros in the amplitudes along the footprint.

\subsection{Angular Fourier modes of the total radio energy along the footprint}
\label{sect:angular_modes}
\begin{figure}
	\includegraphics[width=0.50\textwidth]{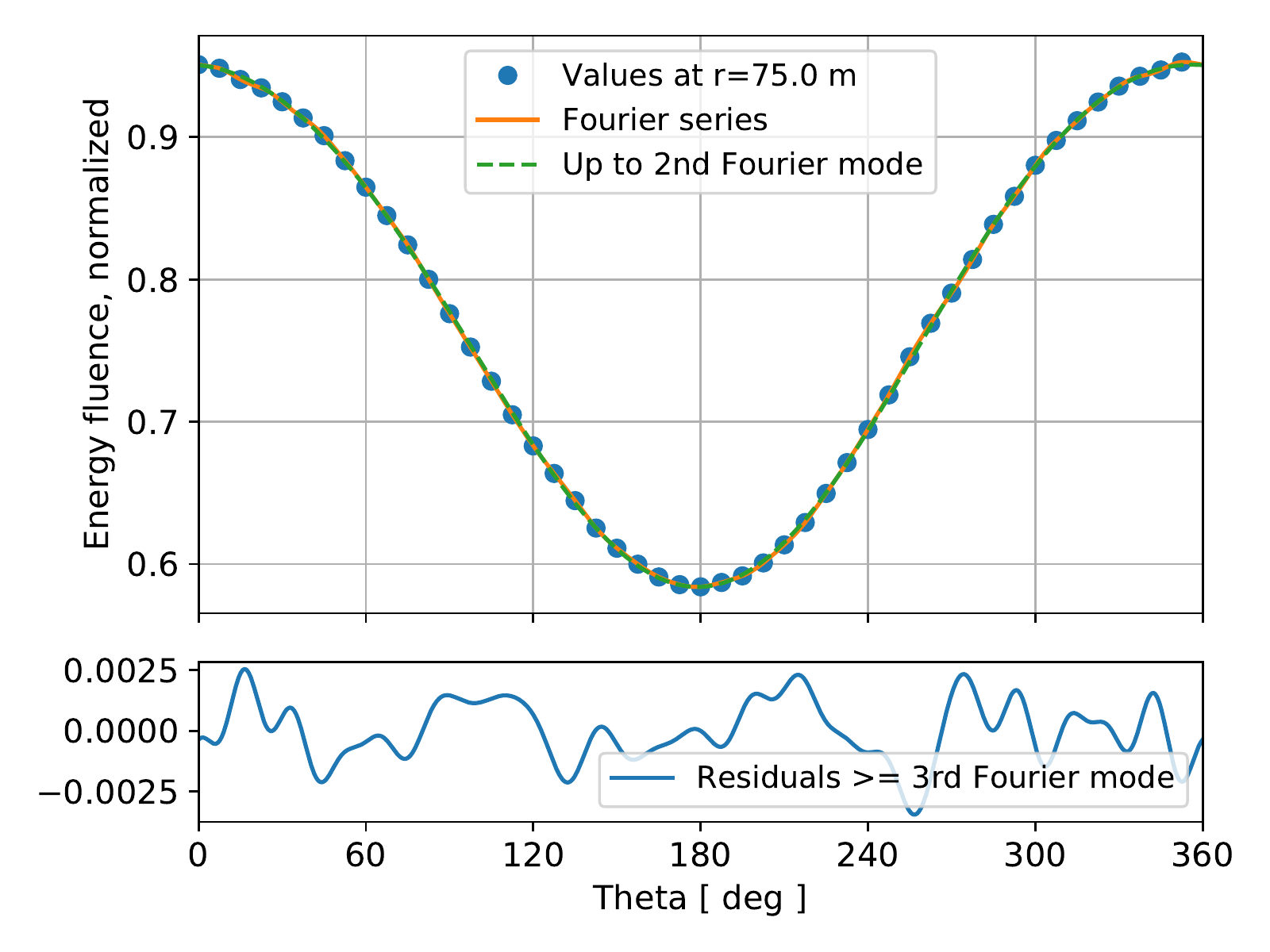}
	\includegraphics[width=0.50\textwidth]{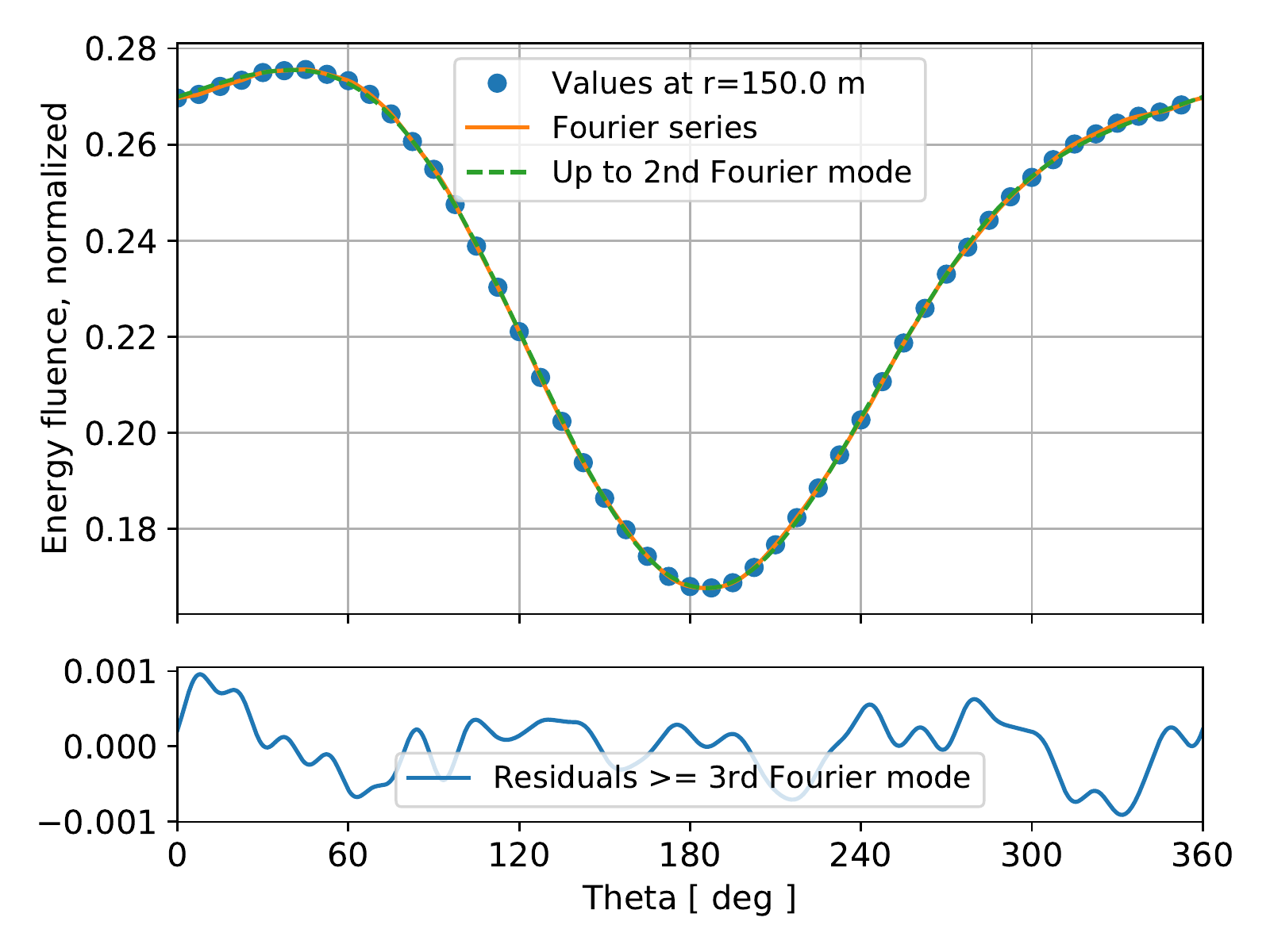}
	\caption{The angular dependence of the pulse energy in an example shower, at $r=\unit[75]{m}$ (left) and at $\unit[150]{m}$ (right). The residuals show that the signal is well represented by Fourier modes up to $2\,\theta$.}
	\label{fig:angular_dependence}
\end{figure}

We demonstrate the suitability of the Fourier modes for this interpolation, using a dedicated simulation with antennas along a circle in the footprint. 
The angular dependence along a circle, at radius 75 and $\unit[150]{m}$ respectively, is shown in Fig.~\ref{fig:angular_dependence}.

The dense sampling, having 48 antennas along a circle, allows us to test the assumption that only the lowest-order Fourier modes are important. 
If this were not the case, simulations would require more sampling (antennas) in the angular direction for any interpolation scheme, until all relevant spatial frequencies are covered.
From the residuals in Fig.~\ref{fig:angular_dependence}, it is clear that errors from a cutoff beyond mode $2\,\theta$ are less than $\unit[0.3]{\%}$ (at thinning level $10^{-6}$). 
This should be sufficient for most purposes, in particular as at this level of detail one should be aware of possible discretization artifacts in the simulated signals, such as from particle thinning.
Thus, the radial grid with 8 points along a circle, typically used in radio simulations of air showers, is sufficient to represent the full pulse parameters. 

\begin{figure}
	\includegraphics[width=0.49\textwidth]{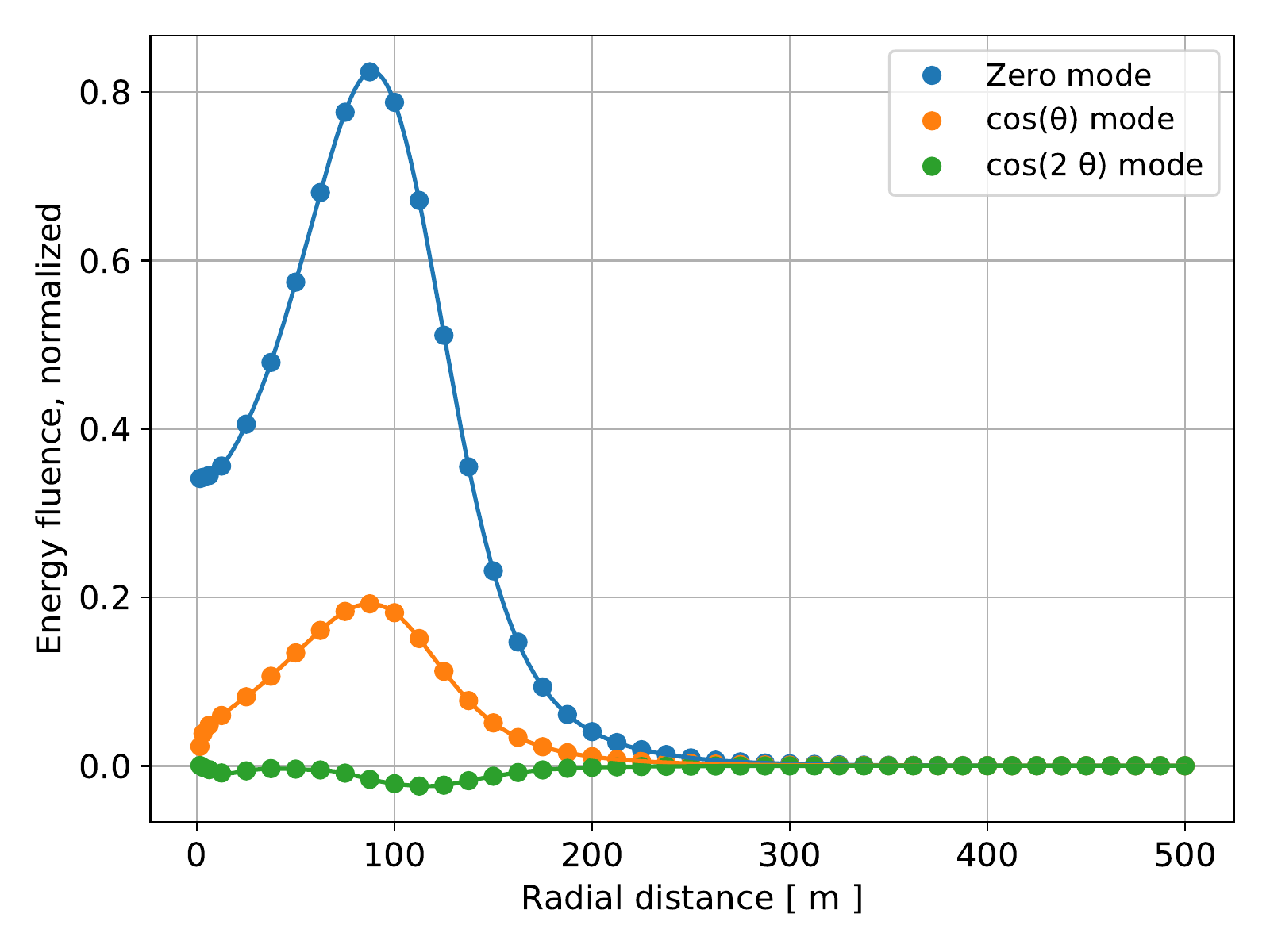}
	\includegraphics[width=0.49\textwidth]{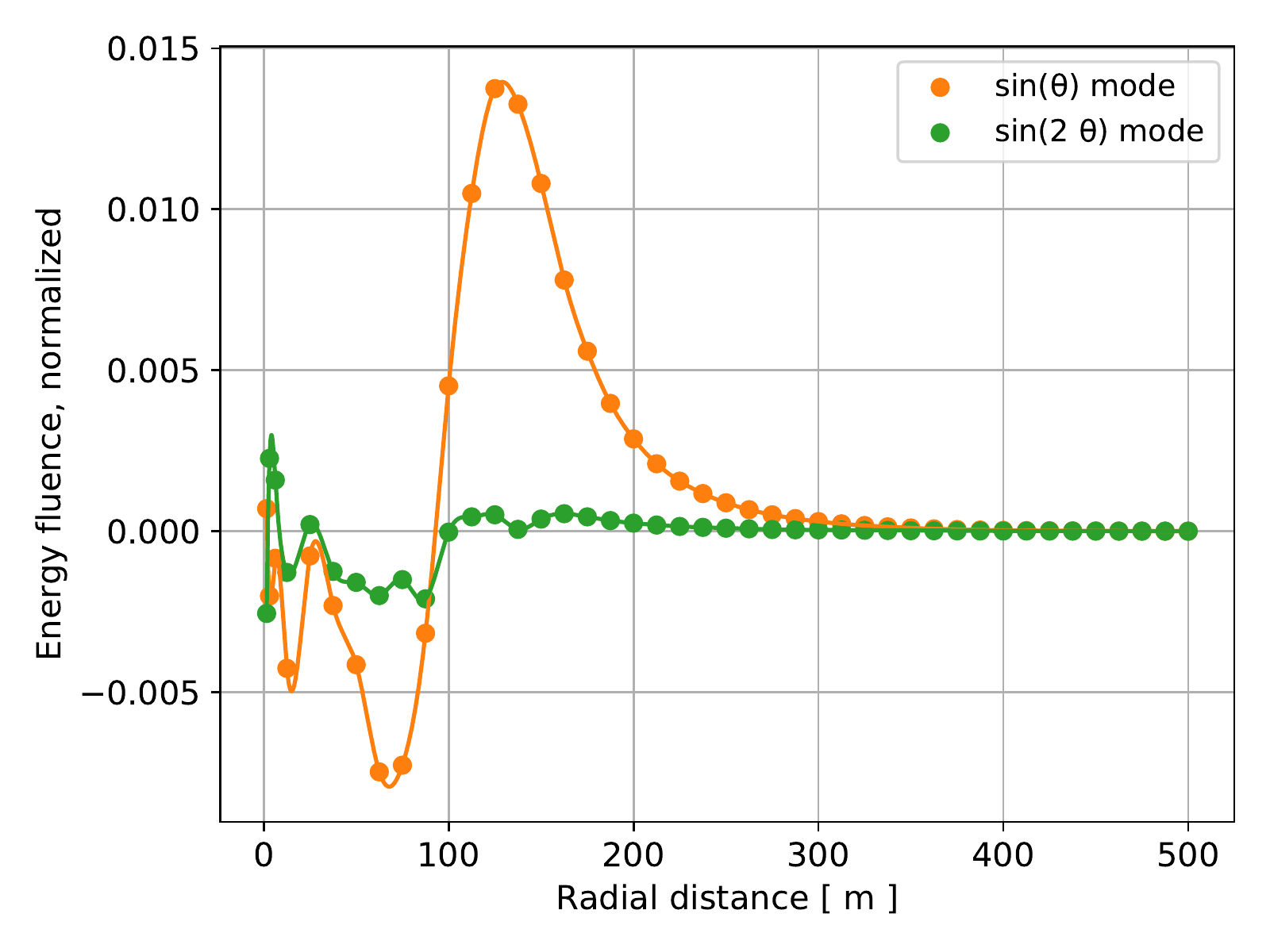}
	\caption{The Fourier modes up to $2\,\theta$ for the energy fluence from all polarizations of an example shower. {\it Left}: the zero mode (circular symmetric) and two cosine modes. {\it Right}: the first two sine modes. Note the vastly different scales on the vertical axes.}
	\label{fig:radial_dependence}
\end{figure}

The radial dependence of the lowest Fourier modes, up to $2\,\theta$, is shown in Fig.~\ref{fig:radial_dependence}.
It shows that the pulse energy has a strong circular-symmetric component (constant or `zero mode' as in $\cos(0\,\theta)$), as well as a secondary component proportional to $\cos (\theta)$, as expected from the geometric and charge-excess contributions, respectively. The contribution proportional to $\cos (2\theta)$ is marginal. 
The sine modes start at $\sin (\theta)$ (no zero mode exists), and its contribution is quite small, see the scale difference in Fig.~\ref{fig:radial_dependence}. Higher-order modes are negligible.

It should be noted that the method only interpolates between the minimum and maximum radius where antennas have been simulated. On a radial grid this means an inner circle around the core is not covered; in this case the inner $\unit[1.6]{m}$. The exact behavior of the signals very near the shower core is not fully understood; while an extrapolation to $r=0$ is possible as the non-zero modes should vanish there, this is not pursued at present as it has little added value.

\subsection{Accuracy of interpolation of the total radio energy fluence and pulse amplitude}\label{sect:results_energy}

We have evaluated the energy fluence over all polarizations, and used the single-value interpolation from Sect.~\ref{sect:energy_interpolation} to obtain the estimated values at the 250 random antenna positions we simulated.
This is done for all showers in the ensemble as described in Sect.~\ref{sect:parameters}.
We use a 30 to $\unit[500]{MHz}$ and a $30$ to $\unit[80]{MHz}$ frequency band to calculate the pulse energy and amplitudes.

\begin{figure}
	\includegraphics[width=0.50\textwidth]{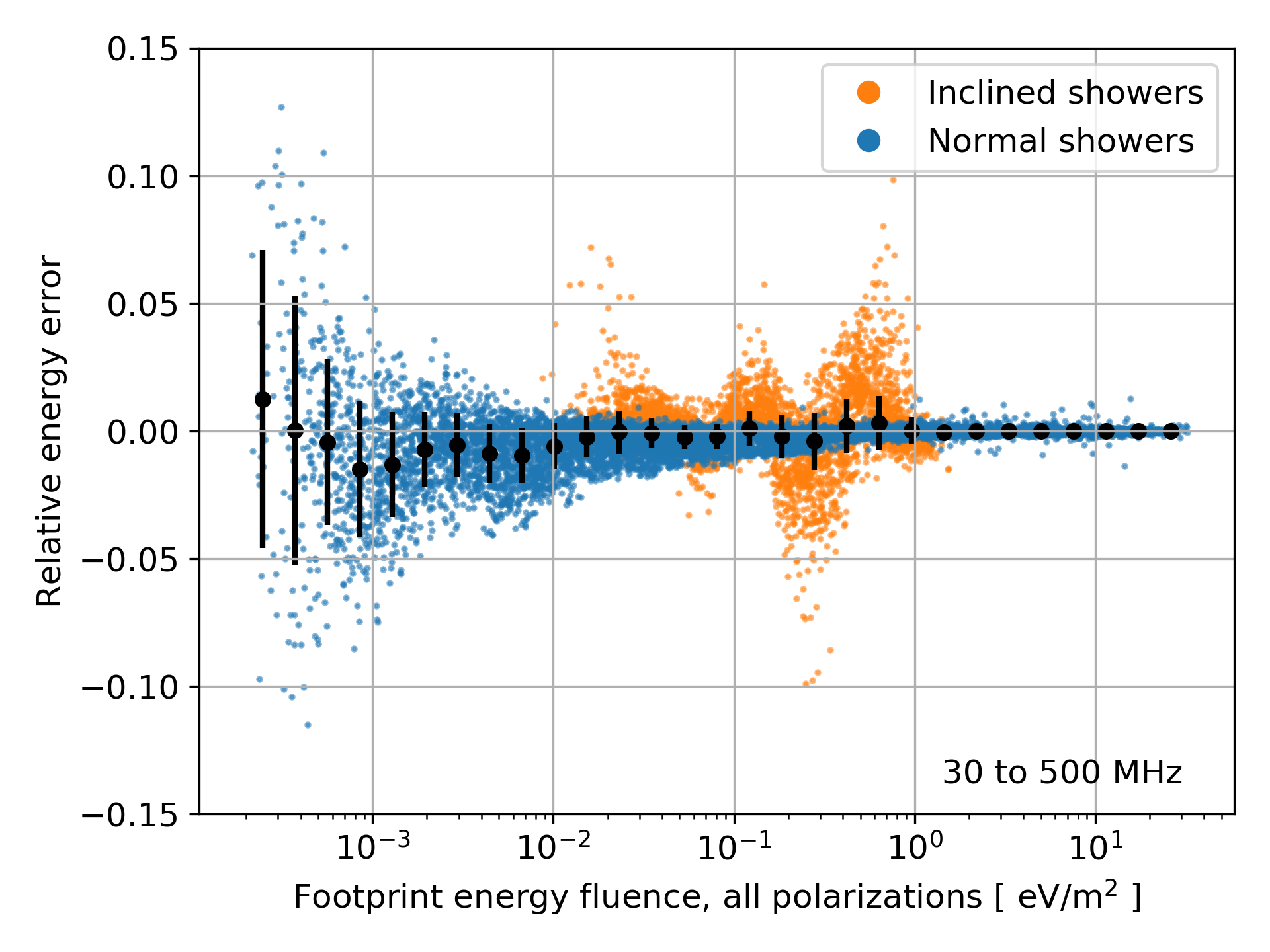}
	\includegraphics[width=0.50\textwidth]{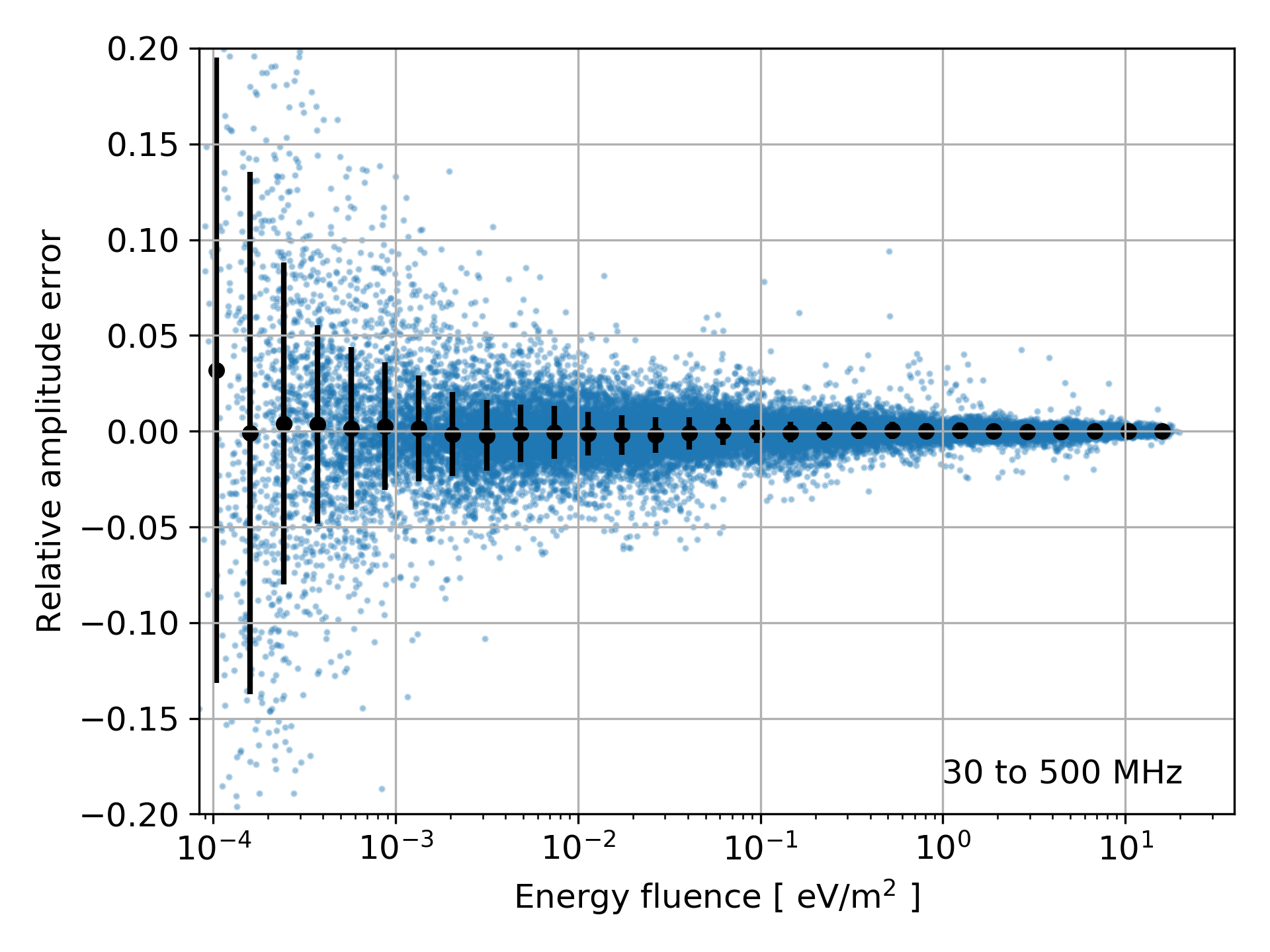}
 	\includegraphics[width=0.50\textwidth]{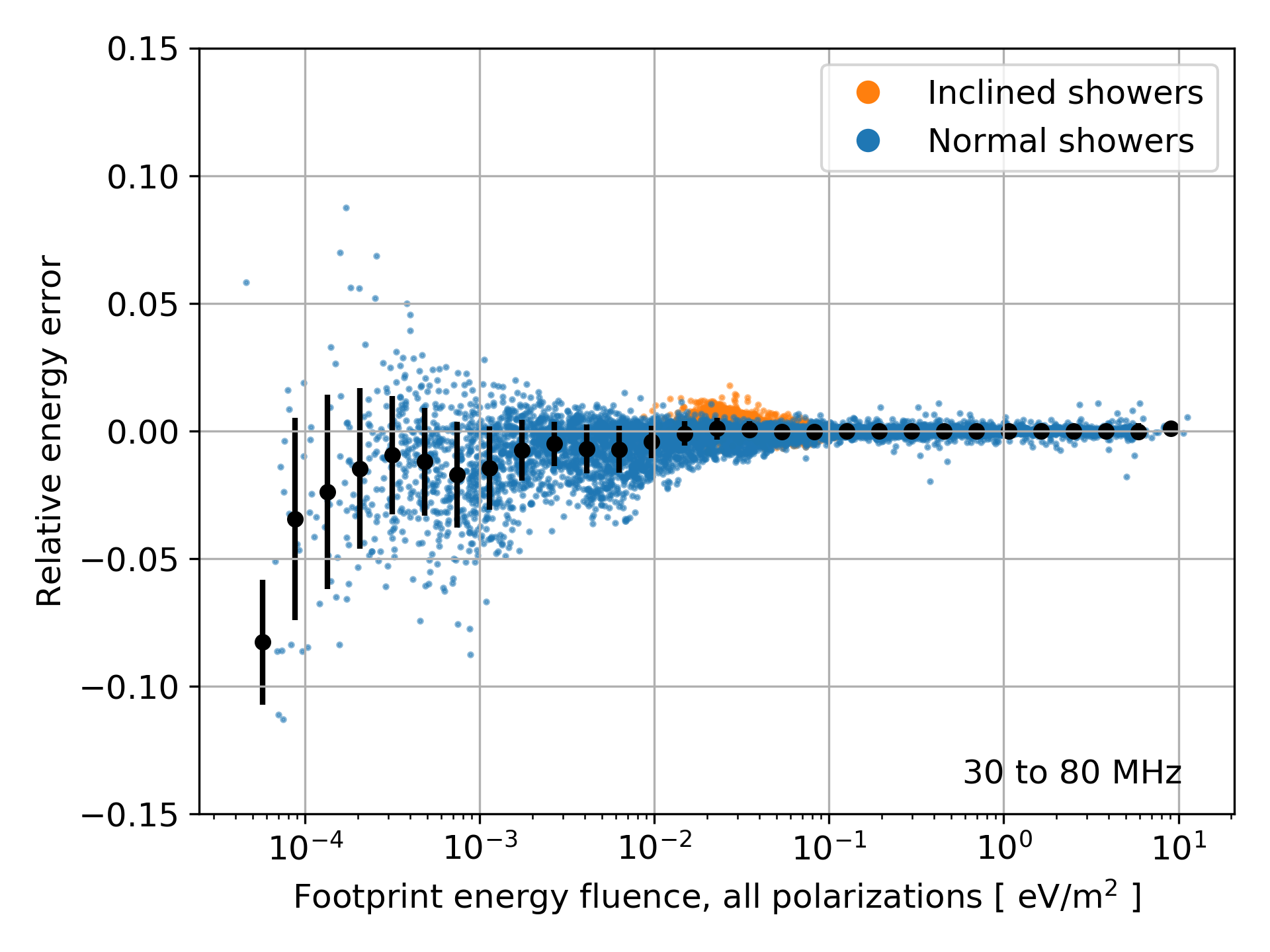}
	\includegraphics[width=0.50\textwidth]{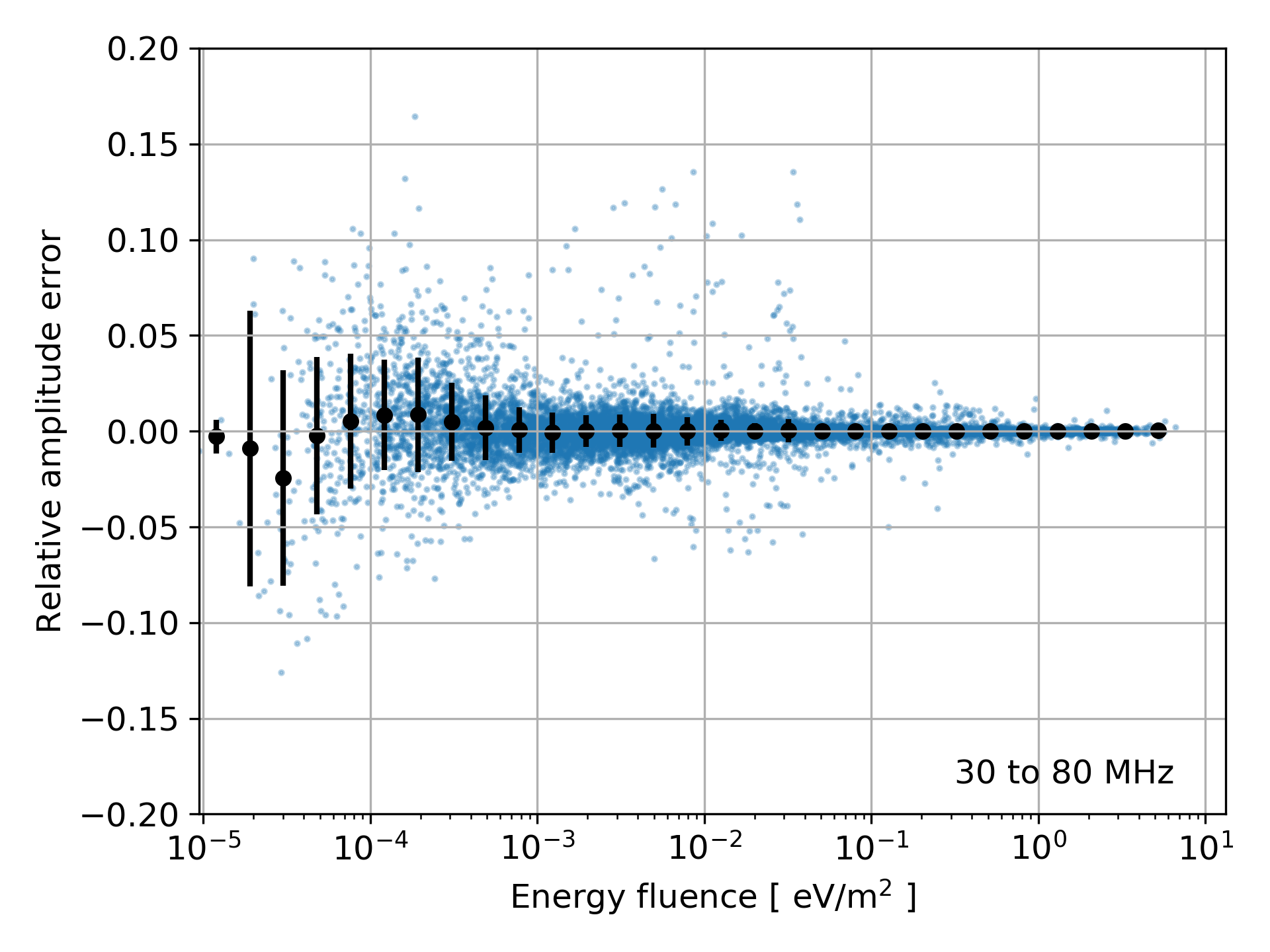}
	\caption{Left column: the relative interpolation error in the total signal energy fluence. Each dot represents a test antenna in one shower of the test ensemble. Black dots show the binned mean, and their error bars show the standard deviation. Right column: the same for the relative interpolation error in the pulse amplitude, per polarization. Top row: Frequency range of 30 to $\unit[500]{MHz}$. Bottom row: Frequency range of 30 to $\unit[80]{MHz}$}
	\label{fig:relative_errors_single}
\end{figure}

In the left panels of Fig.~\ref{fig:relative_errors_single} we show the relative error in the interpolated values, versus the true values, for all showers in the ensemble.
The typical relative error, given by the standard deviation in these plots, varies from around $\unit[1]{\%}$ at low fluence levels of e.g.~$10^{-3}$ to $\unit[10^{-2}]{eV/m^2}$, to about $\unit[0.1]{\%}$ above $\unit[1]{eV/m^2}$. Note that when simulating higher primary energies, these fluence numbers are expected to shift roughly proportionally, as they also depend on the simulated geometry.
For the $30$ to $\unit[80]{MHz}$ band, the numbers are slightly better still.
At low fluence levels, in this case below $\unit[10^{-2}]{eV/m^2}$, there appears to be a small negative bias around $\unit[1]{\%}$.
A difference between inclined showers (here at $80^{\circ}$ zenith angle) and `normal' showers is apparent here; for zenith angles up to $45^{\circ}$ the interpolation is more accurate. This is most likely for geometric reasons; the inclined showers at $80^{\circ}$ zenith angle feature a thin Cherenkov ring near $\unit[750]{m}$ from the core. The corresponding large derivatives, also beyond this distance, would require somewhat more radial sampling locally.

Another typical use case is interpolating the amplitude of the pulse, i.e., the absolute-maximum value in each polarization, along the footprint.
The results are shown in the right panels of Fig.~\ref{fig:relative_errors_single}, showing relative amplitude error versus energy fluence. 
Naturally, the relative error grows towards lower energy levels, and gradually reduces from about $1$ to $\unit[0.1]{\%}$ over the highest three orders of magnitude in fluence. In this range there are also outliers up to about $\unit[5]{\%}$, as visible in the graph, but these are a small minority of cases. 
The deviations in inclined showers that are present in the fluence plot, are not visible here; as the amplitude is roughly proportional to the square root of the fluence, the derivatives along the footprint are smaller.

For the $30$ to $\unit[80]{MHz}$ band, the amplitude is more easily represented, and typical errors are $\unit[0.5]{\%}$, dropping to $\unit[0.1]{\%}$ at high fluence, and again increasing for the weakest signals.
Bias is negligible in either case.

\subsection{Accuracy of interpolation of the pulse time series}
\label{sect:results_pulse}

We have interpolated the full pulse time series for the two `on-sky' polarizations from the CoREAS output, and compare them to the true pulse signals at the 250 test locations per shower. We have used the method based on interpolating phasors, or `method (1)' as described in Sect.~\ref{sect:method_pulse}; the method based on pulse timing gives very similar results for the test ensemble used here.
Note that the polarization directions were intentionally chosen {\emph {not}} to align with $\mathbf{v}\times\mathbf{B}$ and $\mathbf{v}\times(\mathbf{v}\times\mathbf{B})$, to avoid having (near-)zero amplitudes along circles in the footprint. When needed, they can be rotated to that frame after interpolation.
As metrics for the accuracy of the interpolation, we use a normalized cross-correlation, an amplitude comparison, and an energy fluence comparison between the interpolated and true signal at the test locations.

\subsubsection{Normalized cross-correlation as a metric for the agreement in pulse shape}

The normalized cross-correlation in the time domain is a number between -1 and +1, akin to the Pearson correlation coefficient.
It is defined, for signals $f(t)$ and $g(t)$ as
\begin{equation}\label{eq:crosscorrelation}
(f \times g)(\tau) = \frac{\int_0^{t_{\rm max}} f(t)\, g(t+\tau)\, dt}{\sqrt{\left(\int_0^{t_{\rm max}} f(t)^2\, dt\right)} \sqrt{\left(\int_0^{t_{\rm max}} g(t)^2\, dt\right)}}.
\end{equation}

Generally the numerator is computed using a multiplication in the Fourier domain, followed by an inverse FFT.
After normalization it is insensitive to scale factors (like amplitude errors), which are therefore treated separately. 
Values should be near unity for an accurate reconstruction.

We evaluate this correlation at $\tau=0$, so we also test the arrival time matching between reconstructed and true signals.
Another option, if one does not care about the exact timing, is to optimize over $\tau$ and find the largest positive value of the cross-correlation. This value will be, by definition, as least as high as the values presented below.
We also use this to note the error in pulse arrival time in the interpolated traces, as defined from the optimized value of $\tau$.

This metric is a useful choice, also when considering for example interferometry or beamforming use-cases where cross-correlations are fundamental ingredients. 
We take the cross-correlation between the true and the interpolated pulse, in a frequency bandwidth from $30$ to $\unit[500]{MHz}$ or from $30$ to $\unit[80]{MHz}$. 

\begin{figure}
	\includegraphics[width=0.50\textwidth]{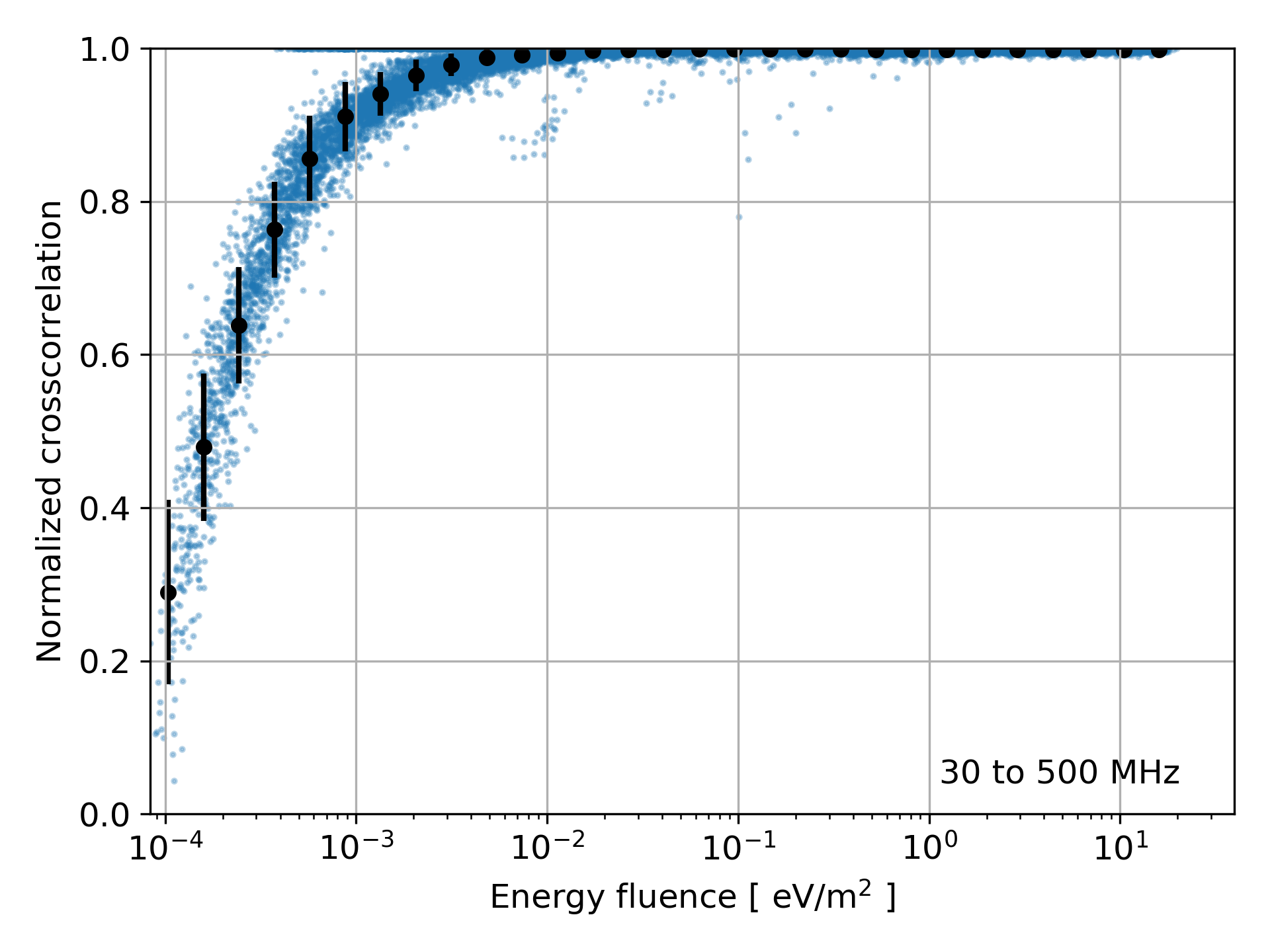}
	\includegraphics[width=0.50\textwidth]{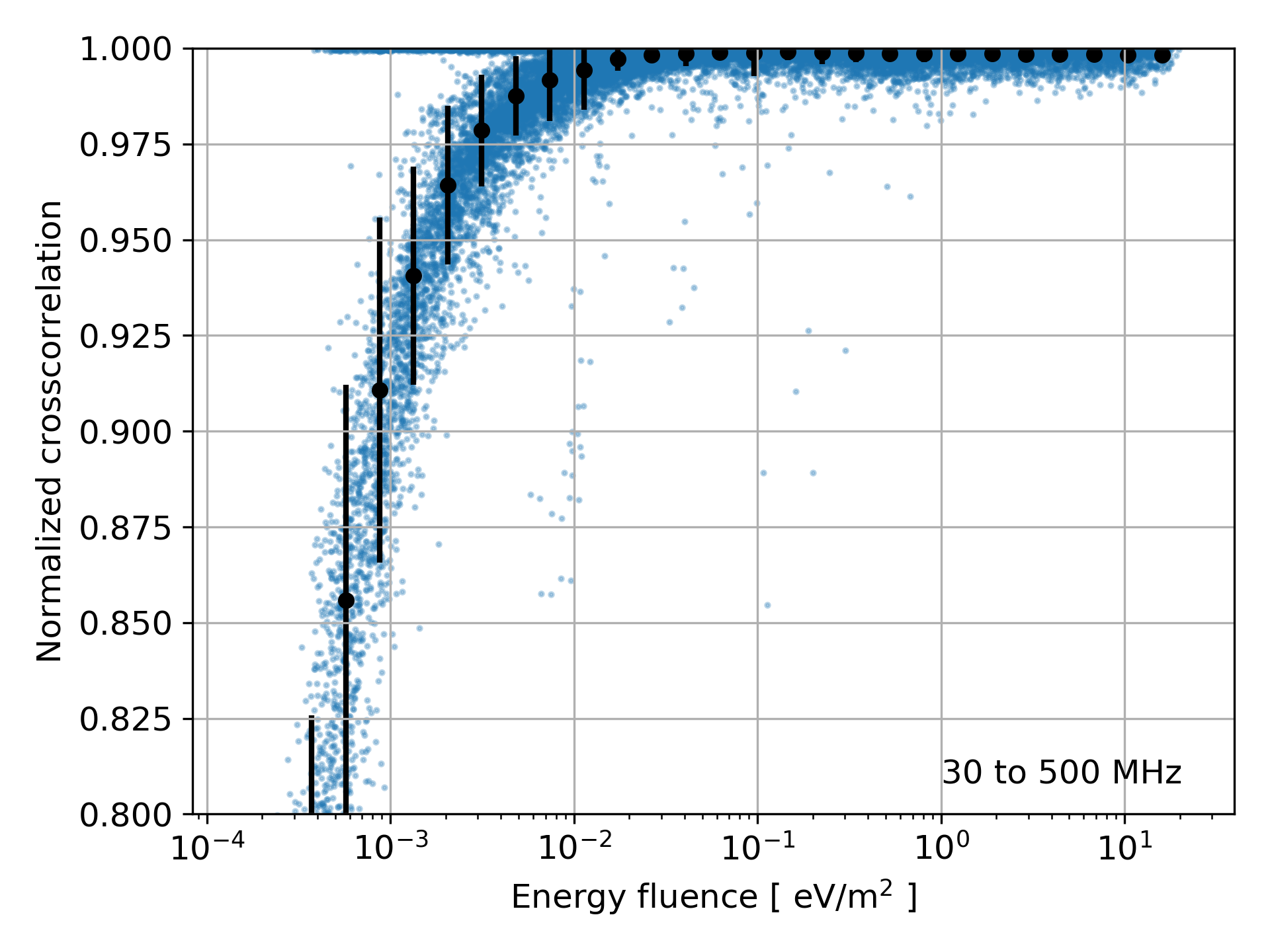}
	\caption{Left: Cross-correlation of interpolated and true pulse at the random test positions in the footprint, for all showers with geomagnetic angle above 10 degrees. Frequency range is 30 to $\unit[500]{MHz}$. Right: zoomed in to the higher cross-correlation values.} 
	\label{fig:CC_results_500}
\end{figure}

In Fig.~\ref{fig:CC_results_500}, we show the cross-correlation results for our simulated ensemble, plotted versus energy fluence.
We have selected only showers with geomagnetic angle above 15 degrees, as showers with very low geomagnetic angle give considerably poorer reconstructions. When the geomagnetic component is very weak or absent, signals may have near-zero amplitude for some positions along a circle in the footprint, leading to problems in the Fourier interpolation of phases (not on absolute amplitudes themselves).
In practice, very few observed showers will have such a small geomagnetic angle, as it covers only a small area on the sky, and the weak signals make them poorly detectable.

Typical values are around 0.998, with a small minority of values going below 0.990. There are about 20 low-outliers above about $\unit[3]{eV/m^2}$ in fluence; these are not there when using the timing-based method instead of interpolating phasors. However, the amplitude accuracy as shown in the next section, has fewer outliers when using phasors.
Compared to a total of 21000 data points, these are quite rare.

Towards lower fluence levels, the accuracy drops. This arises mainly from the high-frequency content, which decays more rapidly with core distance than the low-frequency signal. The fluence level where this drop sets in depends on both the primary energy and the thinning level setting, as demonstrated in Sect.~\ref{sect:thinning} below.

\begin{figure}
	\includegraphics[width=0.50\textwidth]{{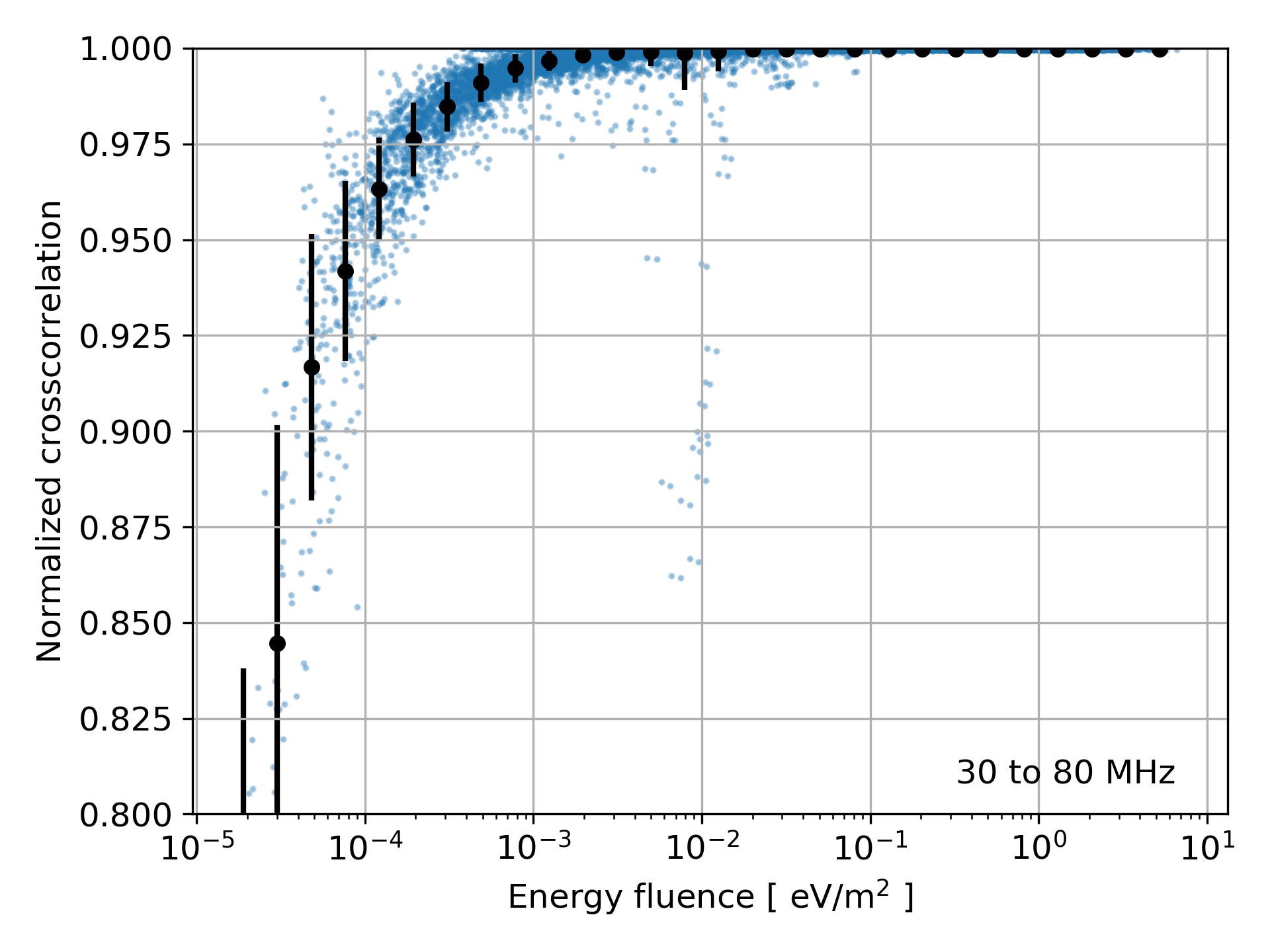}}
	\includegraphics[width=0.50\textwidth]{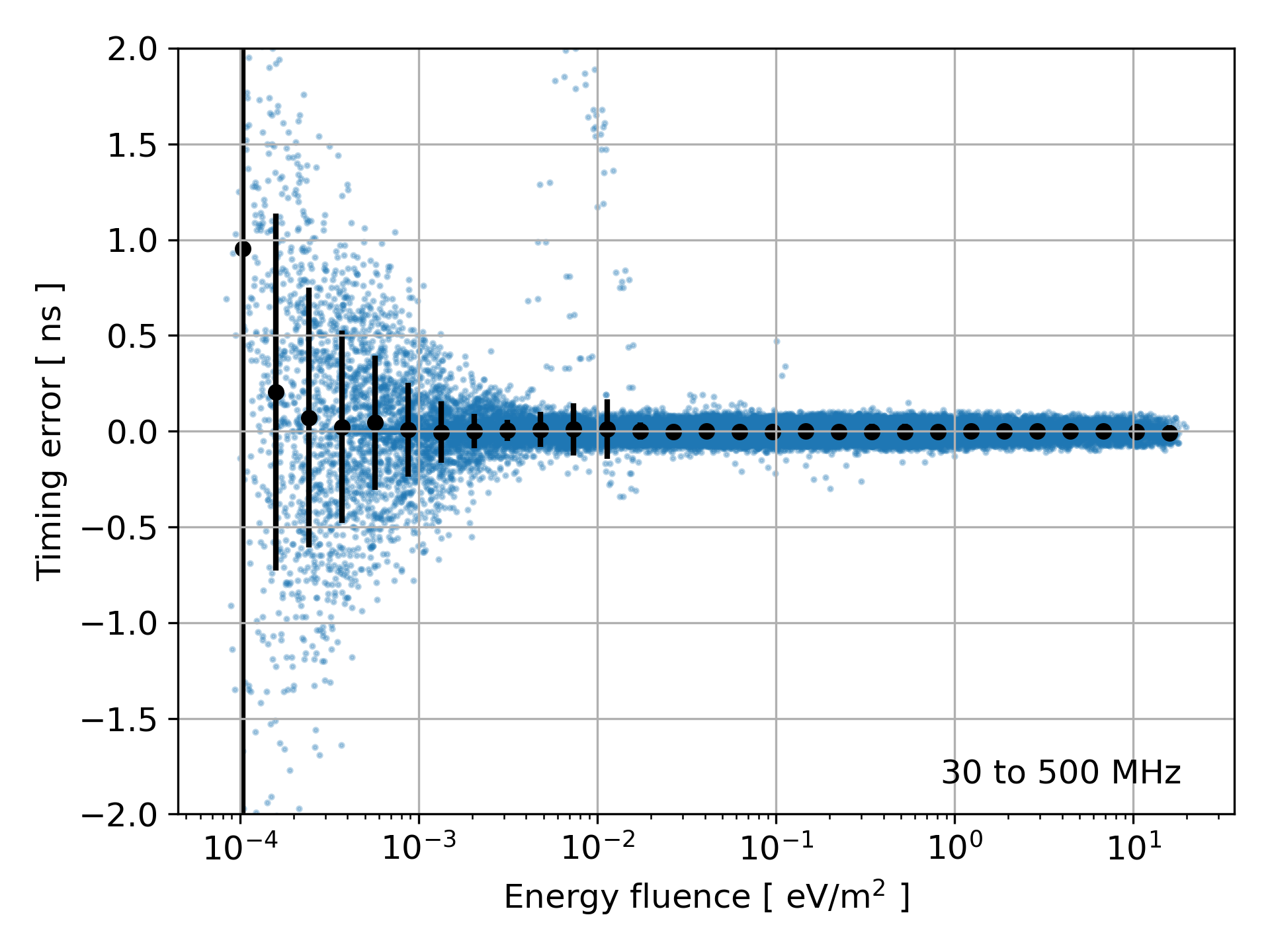}
	\caption{Left: Normalized cross-correlation versus energy fluence per polarization, for the $30$ to $\unit[80]{MHz}$ frequency band. Right: timing accuracy as a function of energy fluence, $30$ to $\unit[500]{MHz}$.}
     
	\label{fig:CC_80_and_timing}
\end{figure}

In the 30 to $\unit[80]{MHz}$ range typically used at e.g.\ LOFAR and AERA, interpolation quality is typically very high, as shown in Fig.~\ref{fig:CC_80_and_timing}; correlation values of $0.999$ are standard at higher pulse energies.
The outliers near $\unit[10^{-2}]{eV/m^2}$ come from a single inclined shower, in a few positions near $\unit[1300]{m}$ from the shower core, which is about twice the radius of the Cherenkov ring. This may require a somewhat denser (radial) sampling with simulated antennas.

Typical timing errors, found from optimizing the cross-correlation by varying $\tau$ in Eq.~\ref{eq:crosscorrelation} are~${\Delta t\sim \unit[0.04]{ns}}$, which should be accurate enough for any interferometric analysis in this frequency range. 
They are plotted in the right panel of Fig.~\ref{fig:CC_80_and_timing}, showing only a few outliers stemming mostly from weak signals.

\subsubsection{Accuracy of pulse amplitude and energy fluence in each polarization}

\begin{figure}
	\includegraphics[width=0.50\textwidth]{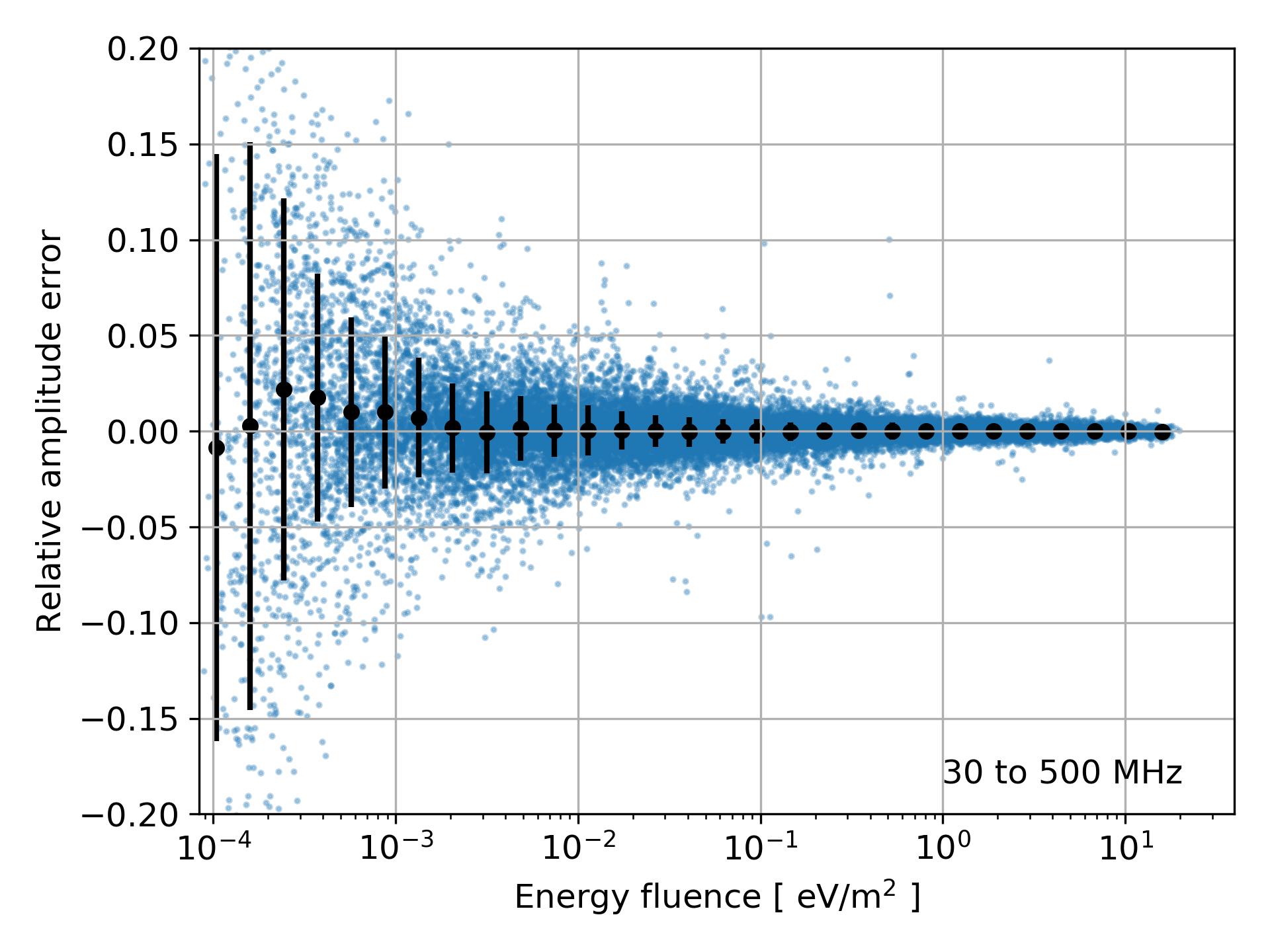}
	\includegraphics[width=0.50\textwidth]{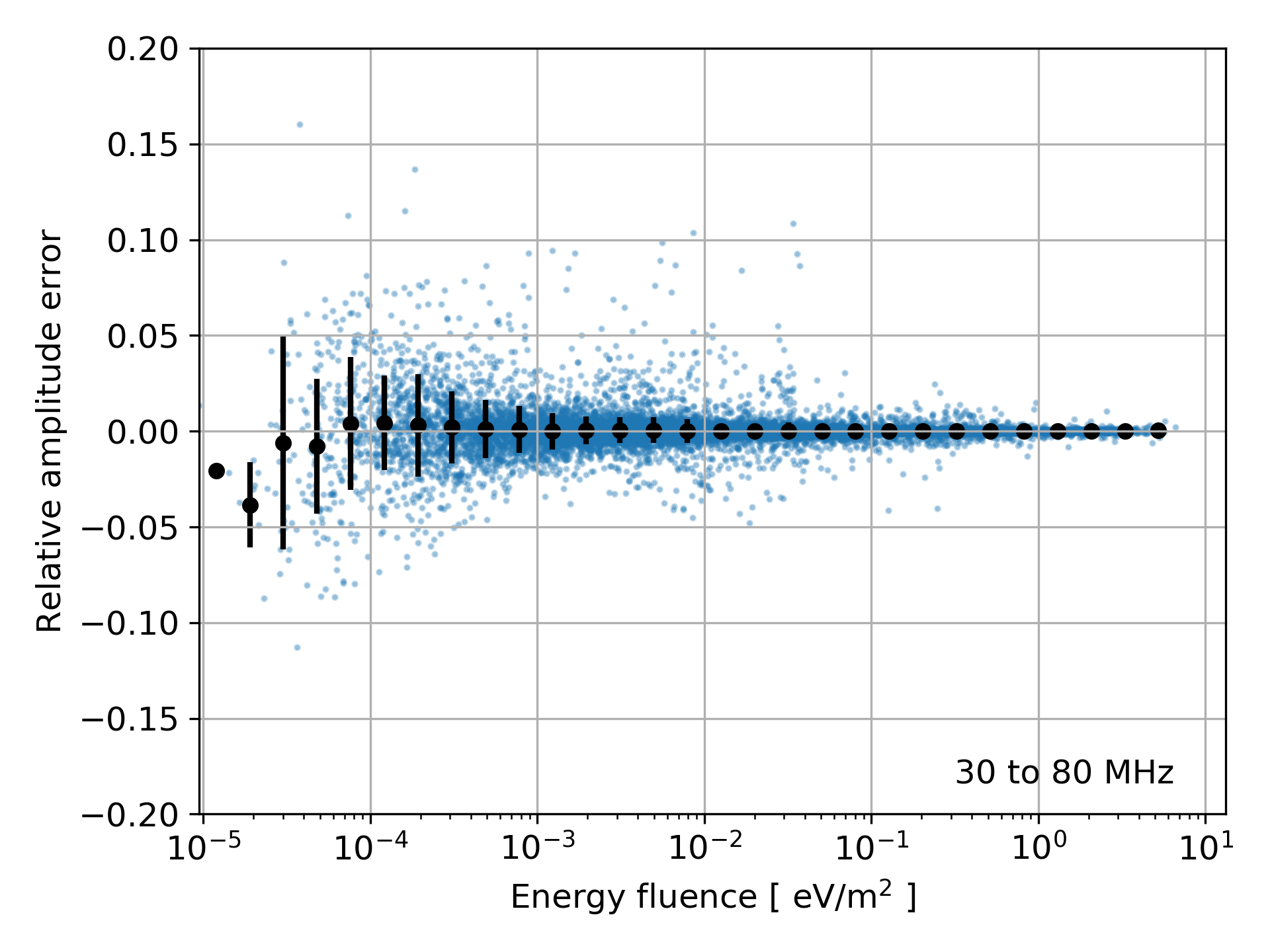}
	\caption{Relative amplitude errors as found in full pulse interpolations. Amplitudes are per polarization; both polarizations are shown together. Left: frequency range 30 to $\unit[500]{MHz}$. Right: range 30 to $\unit[80]{MHz}$.}
	\label{fig:rel_ampli}
\end{figure}

Relative errors in pulse amplitude, versus energy fluence, are shown in Fig.~\ref{fig:rel_ampli}. Here, amplitude and energy fluence refer to a single polarization; each of the two on-sky polarizations is treated separately. The results for the two polarizations have been plotted together; as their direction varies with the shower geometry a separation is not meaningful.

For 30 to $\unit[500]{MHz}$, the standard deviation varies from below 0.5 to $\unit[1]{\%}$ for the higher values of the energy fluence. 
There are only a few outliers up to 5 to $\unit[10]{\%}$, but they are a minority; the plot is in fact saturated with data points within a one-sigma range, due to the large number of test positions. As can be expected, the relative errors in amplitude gradually grow towards very weak signals. 
In the $30$ to $\unit[80]{MHz}$ range, standard deviations are on the order of $0.1$ to $\unit[0.3]{\%}$, again increasing towards weak signals.

\subsubsection{The effect of thinning on the interpolation accuracy}\label{sect:thinning}
\begin{figure}
	\centering
	\includegraphics[width=0.50\textwidth]{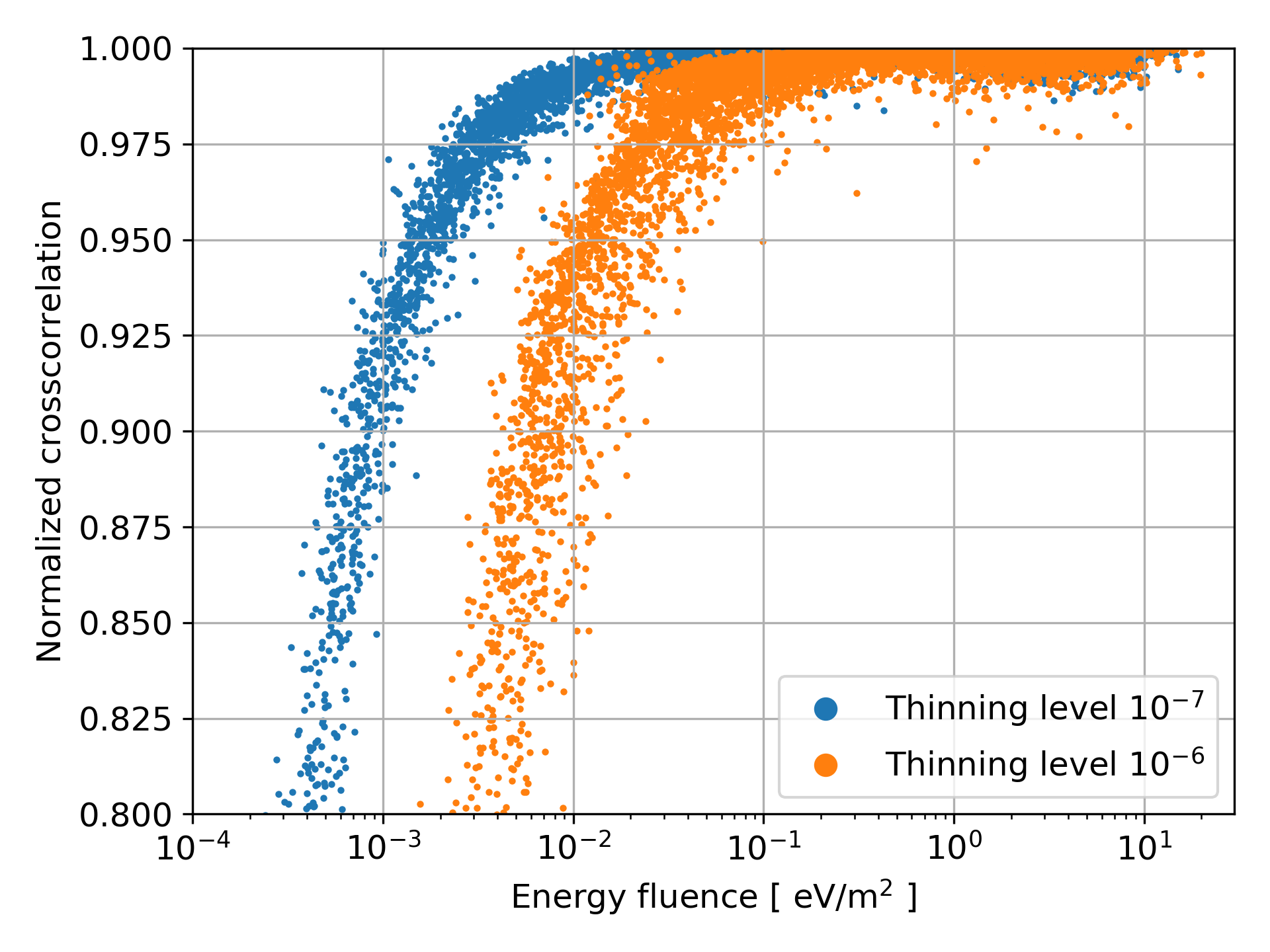}
	\caption{The effect of the thinning level on the cross-correlation between interpolated and simulated pulses. Frequency range is 30 to $\unit[500]{MHz}$. The fluence threshold where degradation sets in will depend on primary energy as well.}
	\label{fig:thinning}
\end{figure}

We have selected 5 shower geometries from the ensemble for which we have run 3 showers each, at the more commonly used thinning level $10^{-6}$ instead of $10^{-7}$ which was used for the analysis above.
The results in terms of normalized cross-correlation, as shown in Fig.~\ref{fig:CC_results_500}, are shown in Fig.~\ref{fig:thinning}.
This clearly demonstrates that the degradation at low fluence levels is due to thinning artifacts rather than intrinsic interpolation mismatch, at least down to a thinning level of $10^{-7}$.
Hence, for high-precision work, setting the thinning level accordingly is important, also in relation to the primary energies being simulated. 
As the thinning level is relative to the primary energy, the fluence level below which the signals degrade will depend on primary energy as well.

\subsubsection{Reliable cutoff frequency and filtering}

\begin{figure}
	\includegraphics[width=0.50\textwidth]{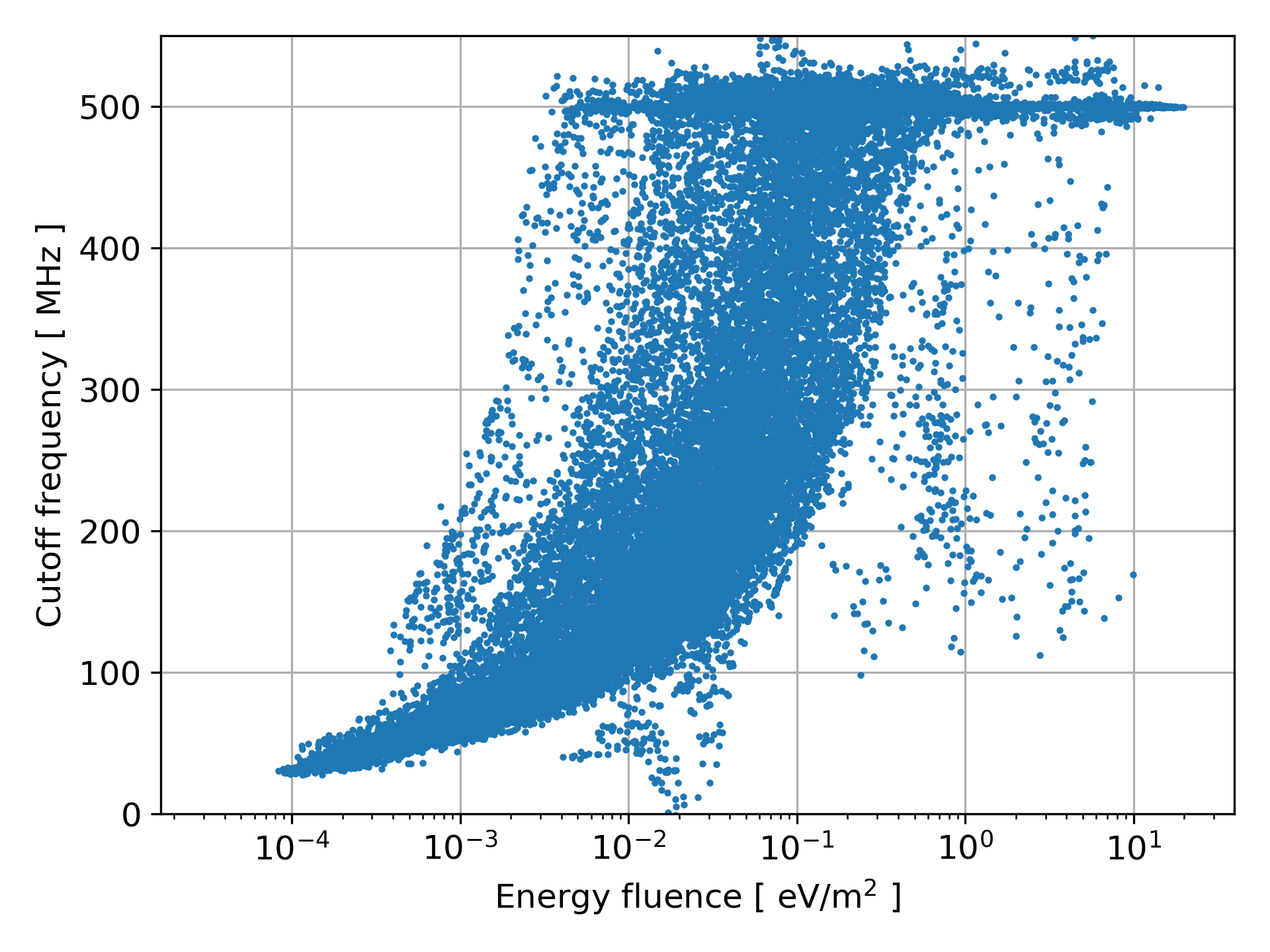}
	\includegraphics[width=0.50\textwidth]{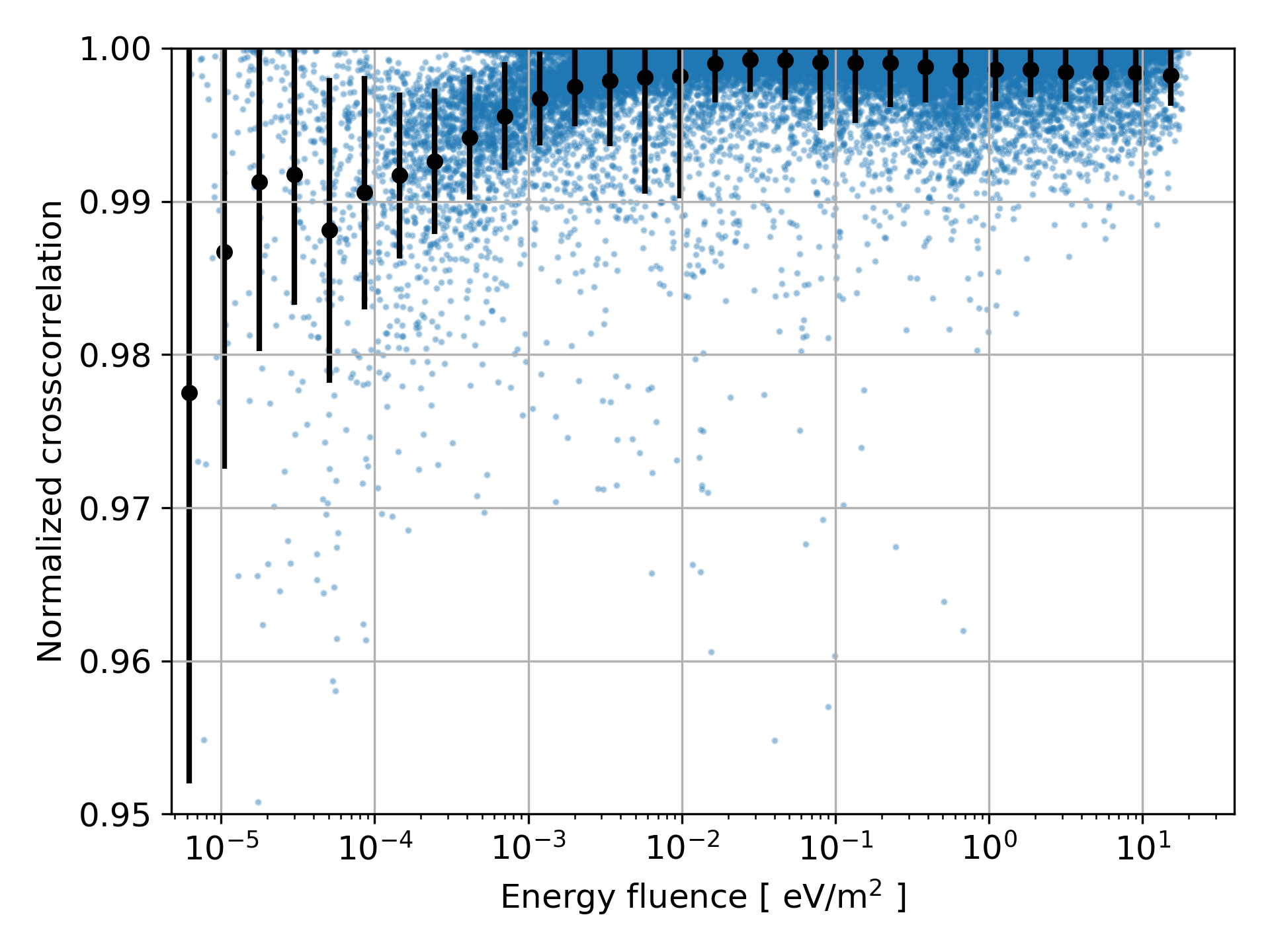}
	\caption{Left: Estimated cutoff frequency as a function of energy fluence. Right: cross-correlation versus energy fluence, range 30 to $\unit[500]{MHz}$, taken from low-pass filtering both the interpolated and original signal up to the local cutoff frequency.}
	\label{fig:cutoff_filtering}
\end{figure}

The method produces an estimated high-frequency cutoff up to which the interpolation is expected to work at full accuracy. This is used internally in `method (2)' as described in Sect.~\ref{sect:method_pulse} but is available in either method.
This estimated frequency cutoff is plotted versus pulse energy fluence in the left panel of Fig.~\ref{fig:cutoff_filtering}, showing a clear falloff towards weaker signals.
The fluence level below which it falls off varies considerably with shower geometry and polarization.
Besides the set of pulse signals that have no meaningful frequency content beyond the cutoff, there are also pulses such as shown in the lower left panel of Fig.~\ref{fig:Pulse_didactic} which have a power spectrum dropping to zero, then rising again. For these, the cutoff frequency will be just below the zero-amplitude frequency.
The low-outliers in the cutoff frequency versus fluence plot are mostly due to these `special' pulses.

When one opts for low-pass filtering the interpolated signals, and comparing them to the original `true' signals with the same filtering applied, one obtains the results as in the right panel of Fig.~\ref{fig:cutoff_filtering}. The main difference with respect to Fig.~\ref{fig:CC_results_500} is that the results now hardly degrade towards weak signals, as the `noisy' high frequency content has been taken out.
This demonstrates that the cutoff frequency is a robust, albeit sometimes overly conservative indicator of the reliably interpolated bandwidth.

\section{Discussion and summary}
We have presented an algorithm to produce the full electric-field time trace as well as pulse amplitude and fluence at any position in a shower radio footprint, given simulated antenna positions on a radial grid in the shower plane.
The interpolation method relies on Fourier representation along circles in the shower plane, making use of the natural geometry of the air shower footprints. Using cubic splines for the radial interpolation makes it a higher-order method everywhere.
Therefore, its performance exceeds what was possible with previously used, linear methods.
One of the most challenging aspects is the interpolation of phases, being periodic quantities. We have found and describe two methods that reliably approximate these in most (about $\unit[99.9]{\%}$ of the) cases.

The pulse prediction quality is measured using normalized cross-correlation, amplitude precision, and timing precision.
We have used an ensemble of 84 proton showers at $\unit[10^{17}]{eV}$ with thinning below $10^{-7}$ of the primary energy, with 250 simulated antennas per shower at random positions, to test the method.
We have analyzed the signals in frequency bands of 30 to 500, and 30 to $\unit[80]{MHz}$, respectively.

Cross-correlation values between directly simulated and interpolated signals are generally around $0.998$ when not in the noise-dominated region, with a few outliers (at the $\unit[0.1]{\%}$ level) down to around $0.90$. 
At lower signal strengths, the signal quality at higher frequencies degrades, and becomes `noise' or artifacts rather than useful signal. The algorithm detects this and estimates a reliable frequency interval and a cutoff frequency. When low-pass filtering up to this cutoff, reconstruction quality remains high even at low signal strengths.

The predicted amplitudes from the interpolation have a standard deviation, depending on energy fluence, of 0.5~to~$\unit[1]{\%}$ with some outliers around $\unit[5]{\%}$, increasing for very weak signals. For the 30 to $\unit[80]{MHz}$ band, the typical error is only $\unit[0.3]{\%}$.
Typical timing errors are very small, around $\unit[0.04]{ns}$, thanks to direct interpolation of the arrival times using the Fourier method.

Together, these results show that the simulation setup with $N=208$ antennas would be sufficient for high-precision work even in the SKA era with ten thousands of antennas per detected shower, including for example interferometric analysis where accurate timing and pulse shape are essential. Even though a relatively low number of antennas is sufficient for the description, measuring air showers with very many antennas would offer unique advantages as the effects of noise can be greatly reduced, and the instrumentation coverage is dense and much more uniform.

Limitations arise when geomagnetic angles, i.e.\ the angle between magnetic field and shower direction, are very small, below about $15^{\circ}$. These showers were omitted from the analysis here as the results were not reliable. However, it is known that these directions are more complicated to analyze and are often excluded. 
An issue to note is that while it may be a natural choice to use polarization orientations that closely align with the $(\mathbf{v}\times\mathbf{B})$ and  $(\mathbf{v}\times\mathbf{v}\times\mathbf{B})$-direction, interpolation may then cause problems, as in these cases signal intensity becomes (near-)zero along circles in the footprint.
Measuring phases or pulse arrival times is in these cases not well possible there, causing missing data. In a higher-order method such as presented here, data needs to be available well beyond nearest neighbors, or specific missing data mitigation measures are needed. While these exist, we have not pursued this here. We have also not opted to fall back to a linear interpolation for weak signals, which would complicate the method, but to take advantage of the higher-order interpolation wherever signals are reasonably strong.
Rotating the reference frame for the polarizations is usually sufficient to produce accurate results, as shown in the previous section; it is typically needed when incoming direction is close to zenith, or to the north-south axis. After interpolation it can be rotated back.

We have not studied performance for extreme circumstances, for instance detections at high altitude, or using a wider variety of inclined showers, where tuning of the presented method might be needed.

Not inherent to this method, limitations in reconstruction quality come from simulation limitations themselves, most notably from particle thinning artifacts.
Attention to the thinning settings is thus needed as the resulting limitations may indeed be reached in practice, especially when doing precision work with many antennas at higher frequencies.

It is noteworthy in itself that the radio signals throughout their footprints allow for interpolation to this level of accuracy, from a radial grid of only 208 antennas. There is a priori no guarantee that the signals vary this smoothly, over sometimes appreciable distances (such as between antennas on a circle with large radius). Thus, we recognize it as a feature of the radio footprints, and it confirms that the polar grid with 8 `arms' as used throughout the radio community based on experience and intuition alone, has been a sufficient and robust choice.

Demonstration code for this method has been made available, so the method can be used by the radio air shower community as add-on to existing simulations \cite{github_nuradio:2023}. 

\begin{appendix} 
\section{Appendix}\label{sect:appendix}
\begin{figure}[h]
	\includegraphics[width=0.50\textwidth]{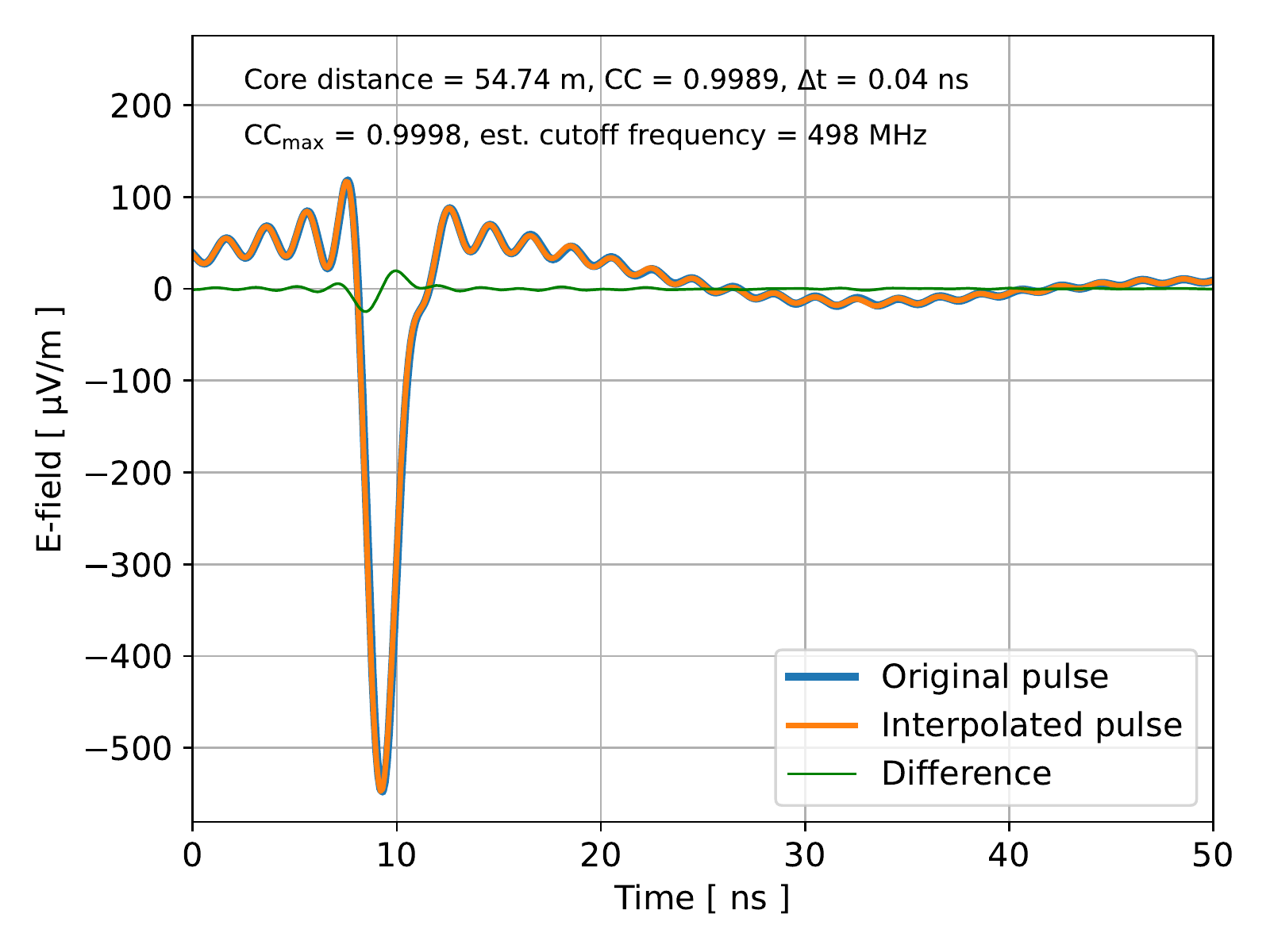}
	\includegraphics[width=0.50\textwidth]{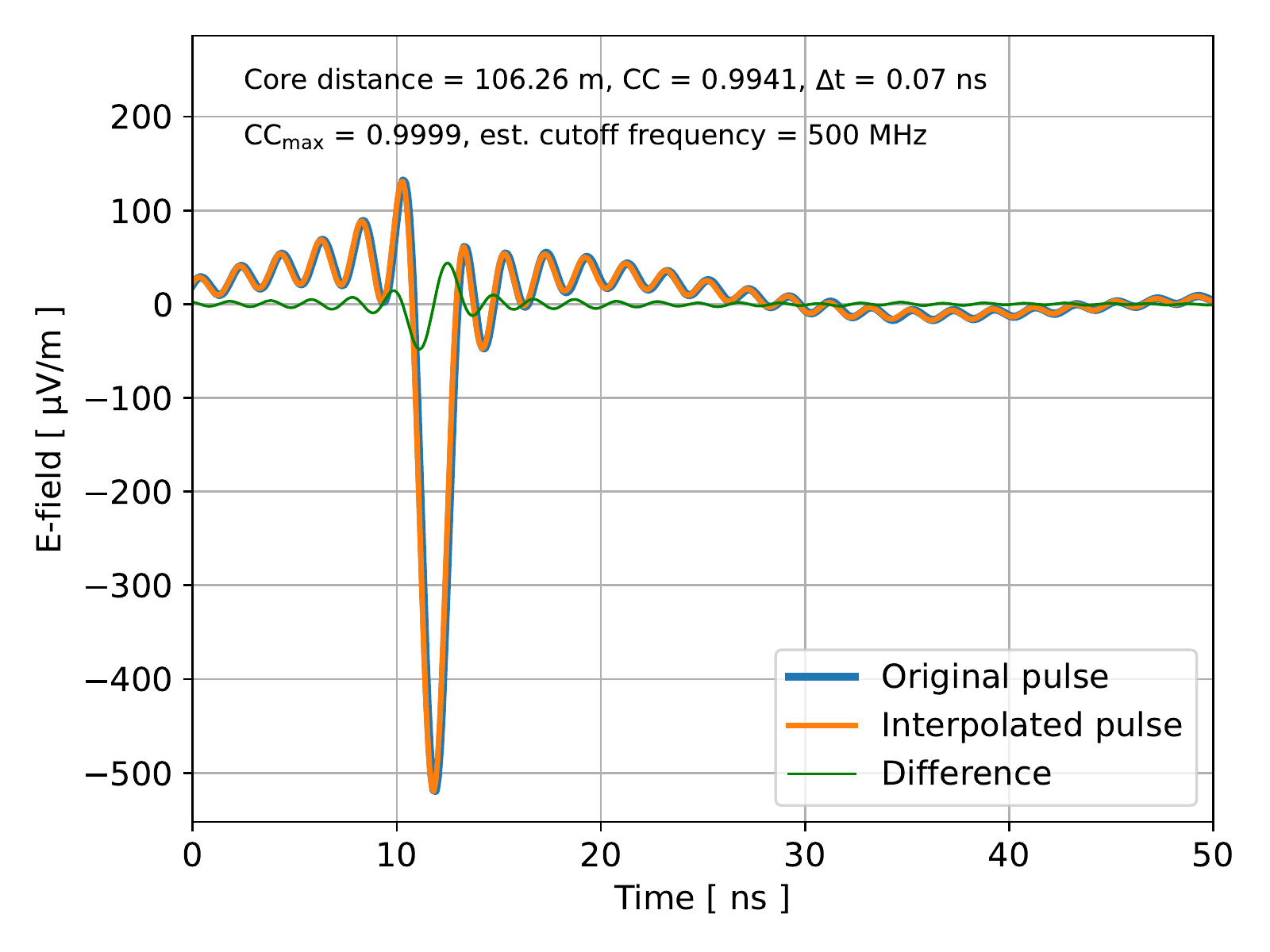}
    \includegraphics[width=0.50\textwidth]{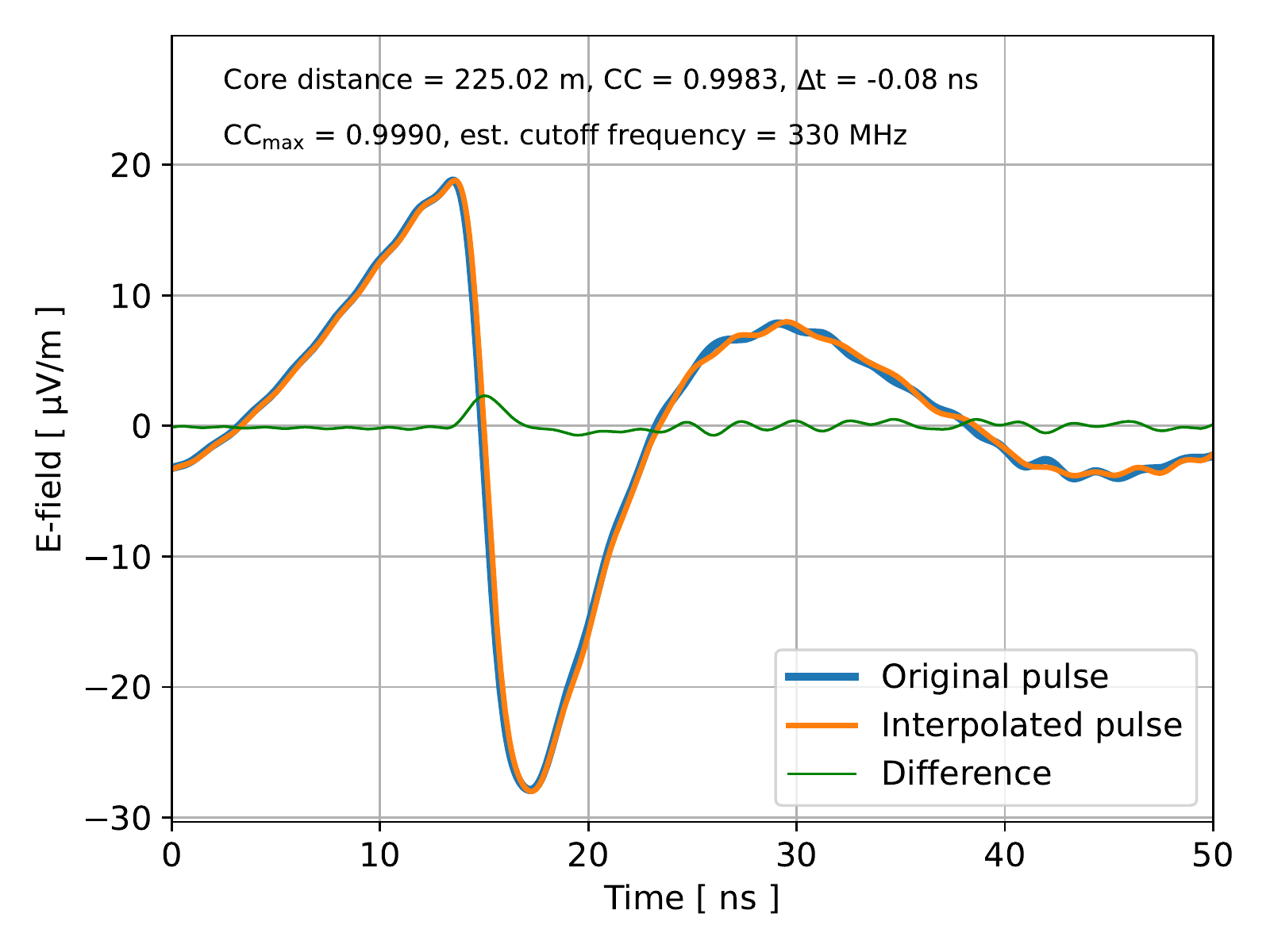}
    \includegraphics[width=0.50\textwidth]{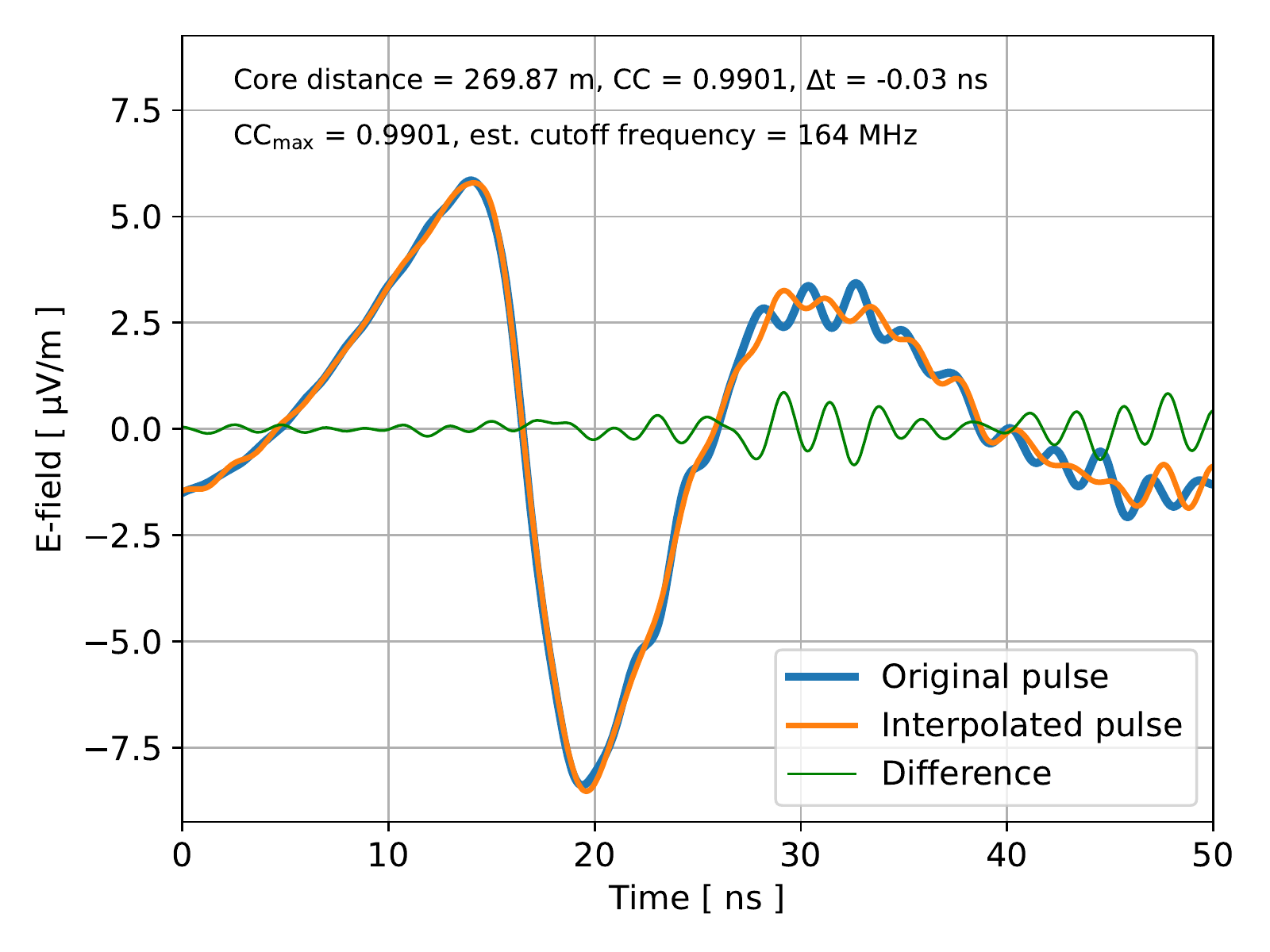}

 \caption{Examples of interpolated pulses compared to the simulated pulses at the same positions.}
	\label{fig:example_reconstruction}
\end{figure}

Here we provide some additional example plots.
In Fig.~\ref{fig:example_reconstruction} the reconstructed and `true' simulated E-field time series are plotted for 4 example locations, at different levels of accuracy. This demonstrates how the given cross-correlation values correspond to accuracy in a time domain plot.
Differences as shown in the residual signal arise from timing mismatches, which are usually very small at $\sim \unit[0.04]{ns}$, and from errors in interpolating the spectra. Both sources of error have been included by evaluating the cross-correlation at a zero time shift.

Indeed, in the upper two plots showing strong pulses inside and near the Cherenkov ring, respectively, the errors arise almost fully from a small timing mismatch. This is seen from the time-optimized crosscorrelation, denoted as $\mathrm{CC}_{\mathrm{max}}$ which is very close to unity.

In the lower two panels, signals further out from the core are shown, which are an order of magnitude weaker. Here, the error is mainly in the high-frequency content of the signal. This is flagged by the method as the estimated cutoff frequency is far below $\unit[500]{MHz}$ here.

\begin{figure}
	\includegraphics[width=0.50\textwidth]{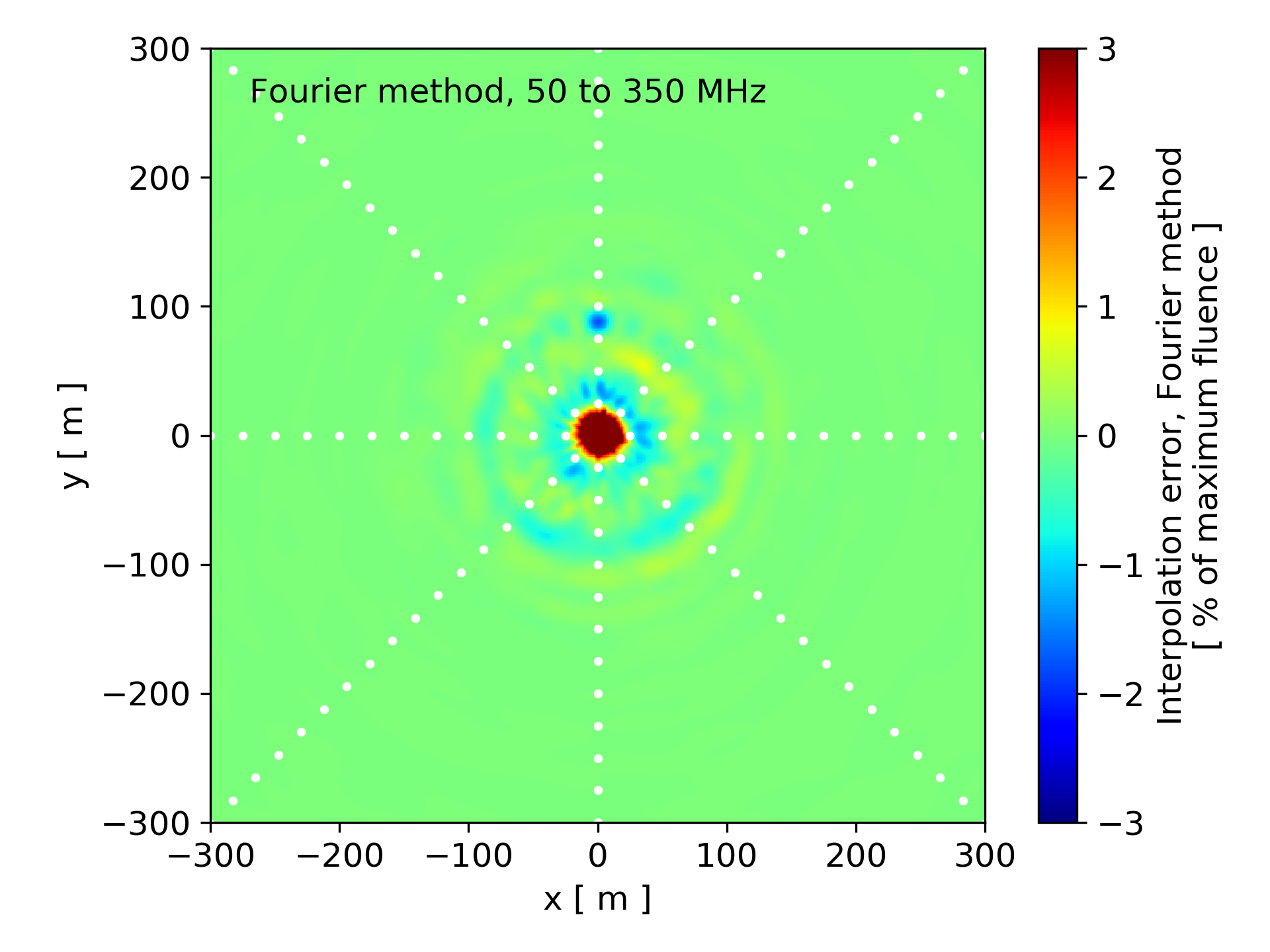}
	\includegraphics[width=0.50\textwidth]{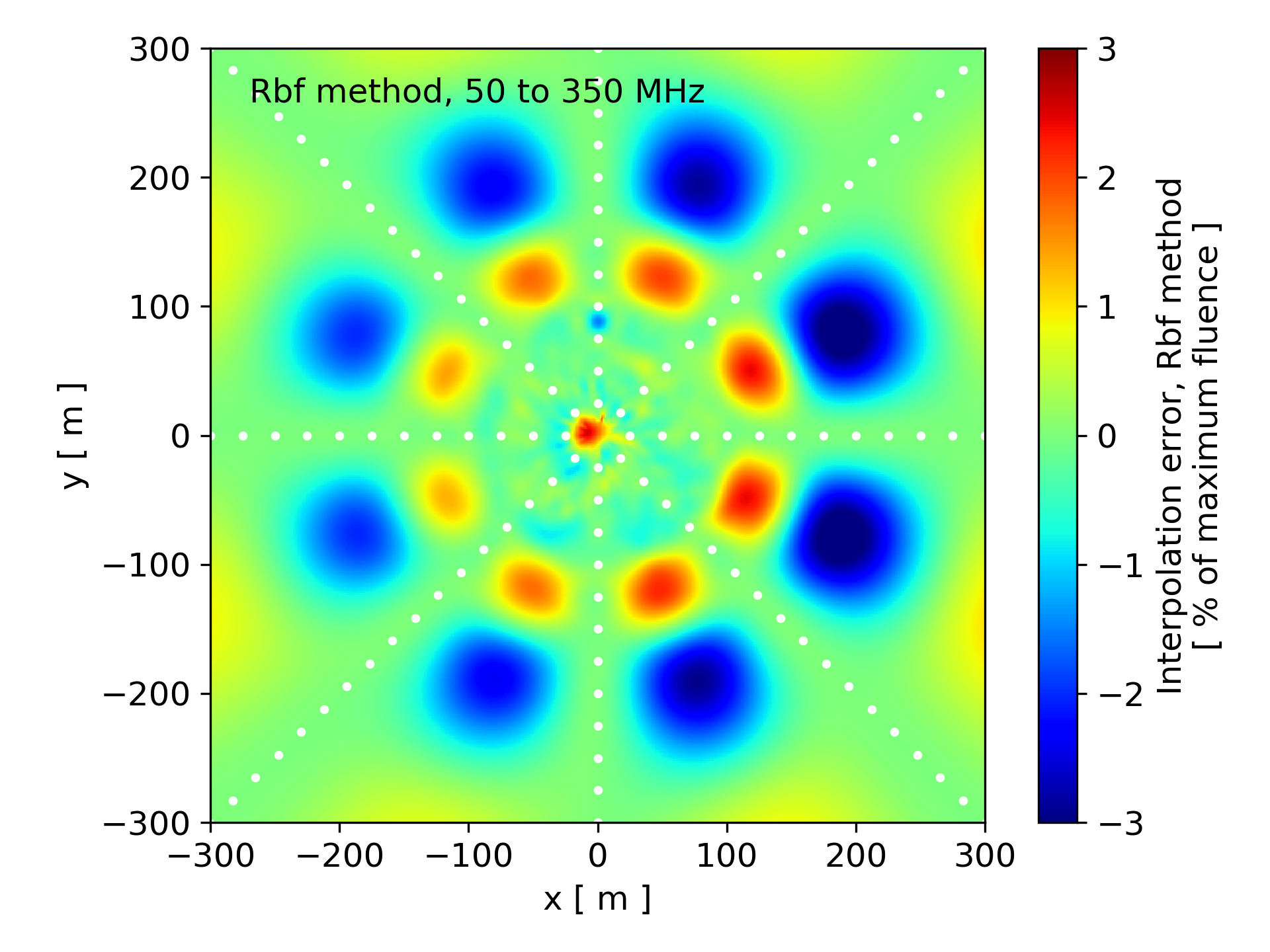}
	\caption{Energy fluence interpolation errors using the Fourier method (left panel) and the previously used method using radial basis functions (denoted Rbf, right panel). Frequency range is $50$ to $\unit[350]{MHz}$, using an equidistant radial grid of 160 antennas.}
	\label{fig:diff_footprints_50_350}
\end{figure}

Another pair of plots is given in Fig.~\ref{fig:diff_footprints_50_350} where we compare the interpolation error of energy fluence along the footprint, between the method presented here and the library method from Scipy we used before.
This was evaluated from simulating a shower with a dense radial grid, which has as a subset the equidistant radial grid with 160 antennas used earlier at LOFAR; the Scipy method required an equidistant grid in the radial direction.
The dense grid has half the radial spacing and 48 `arms' instead of 8, comprising 1920 antennas.
We compare the interpolated footprint using the standard $N=160$ layout with the interpolated footprint using $N=1920$ antennas.
With this large difference in antenna density, this comparison will closely approximate the error with respect to a `true' simulated footprint.

Fig.~\ref{fig:diff_footprints_50_350} shows the difference footprint for the $50$ to $\unit[350]{MHz}$ frequency range relevant for SKA, for the Fourier method presented here, and for Scipy's radial basis function method, respectively. The error is given as a percentage of the maximum fluence. Using the old method (right panel), the errors are relatively large, on the order of $\unit[3]{\%}$ of the maximum fluence (hence even larger when taken relative to the local fluence), in relatively large areas of the footprint. For the LOFAR frequency range this number was below $\unit[1]{\%}$.

Seeing this, we realized that to match the next-level precision to be delivered by SKA, either at least an order of magnitude more antennas would have to be simulated, or a better interpolation method would be needed. This is what motivated the present analysis.

Indeed, the presented method performs quite well in comparison, as seen in the left panel. A few sources of error are still visible, though. First, it is clear that the core region was under-sampled in the $N=160$ layout. And second, the footprints at higher frequencies feature a sharper Cherenkov ring; the higher-order radial derivatives become a source of error there. This is visible as the faint concentric rings in the left panel of Fig.~\ref{fig:diff_footprints_50_350}.

We have addressed both, by adjusting the simulated antenna layout to be non-equidistant radially, as the presented method allows this. Up to $\unit[200]{m}$ from the core, we use half the radial distance, i.e.~\unit[12.5]{m}.
Further out, we increased the radial distance to 25, 50, and $\unit[100]{m}$, respectively.
To cover the inner core region, we have added 3 more radial positions, leaving only the inner $\unit[2]{m}$ uncovered.
This way, the number of antennas to be simulated increases only marginally to 208, and this is the layout we have used throughout the analysis. There may be room for further optimization, but as this depends on details of the use case we leave this for a future analysis when needed.

\end{appendix}
\section*{Acknowledgements}
BMH is supported by ERC Grant agreement No.~101041097; AN and KT acknowledge the Verbundforschung of the German Ministry for Education and Research (BMBF). 
NK acknowledges funding by the Deutsche Forschungsgemeinschaft (DFG, German Research Foundation) – Projektnummer 445154105.
MD is supported by the Flemish Foundation for Scientific Research (FWO-AL991).
ST acknowledges funding from the Abu Dhabi Award for Research Excellence (AARE19-224).

\bibliographystyle{elsarticle-num}
\bibliography{pulse_interpolation}

\end{document}